\documentclass[aps, prl, fleqn, reprint, superscriptaddress, showpacs, twocolumn]{revtex4-1}%
\usepackage{amsmath}
\usepackage{mathrsfs}
\usepackage{amsfonts}
\usepackage{tabularx}
\usepackage{times}
\usepackage{graphicx}
\usepackage{color}
\newcommand{\Pg}{P_{|g\rangle}}
\newcommand{\Pe}{P_{|e\rangle}}
\newcommand{\Pf}{P_{|f\rangle}}
\newcommand{\GG}{{|g\rangle}}
\newcommand{\EE}{{|e\rangle}}
\newcommand{\FF}{{|f\rangle}}
\newcommand{\Asig}{A_{\rm sig}}
\newcommand{\Aref}{A_{\rm ref}}
\newcommand{\Teff}{T_{\rm eff}}

\begin{document}
\title{Thermal and Residual Excited-State Population in a 3D Transmon Qubit}
\author{X. Y. Jin}
\email{jin@mit.edu}
\affiliation{Research Laboratory of Electronics, Massachusetts Institute of Technology, 77 Massachusetts Avenue, Cambridge, Massachusetts 02139}
\author{A. Kamal}
\affiliation{Research Laboratory of Electronics, Massachusetts Institute of Technology, 77 Massachusetts Avenue, Cambridge, Massachusetts 02139}
\author{A. P. Sears}
\affiliation{MIT Lincoln Laboratory, 244 Wood Street, Lexington, Massachusetts 02420, USA}
\author{T. Gudmundsen}
\affiliation{MIT Lincoln Laboratory, 244 Wood Street, Lexington, Massachusetts 02420, USA}
\author{D.~Hover}
\affiliation{MIT Lincoln Laboratory, 244 Wood Street, Lexington, Massachusetts 02420, USA}
\author{J. Miloshi}
\affiliation{MIT Lincoln Laboratory, 244 Wood Street, Lexington, Massachusetts 02420, USA}
\author{R.~Slattery}
\affiliation{MIT Lincoln Laboratory, 244 Wood Street, Lexington, Massachusetts 02420, USA}
\author{F. Yan}
\affiliation{Research Laboratory of Electronics, Massachusetts Institute of Technology, 77 Massachusetts Avenue, Cambridge, Massachusetts 02139}
\author{J. Yoder}
\affiliation{MIT Lincoln Laboratory, 244 Wood Street, Lexington, Massachusetts 02420, USA}
%
%
\author{T. P. Orlando}
\affiliation{Research Laboratory of Electronics, Massachusetts Institute of Technology, 77 Massachusetts Avenue, Cambridge, Massachusetts 02139}
\author{S. Gustavsson}
\affiliation{Research Laboratory of Electronics, Massachusetts Institute of Technology, 77 Massachusetts Avenue, Cambridge, Massachusetts 02139}
\author{W. D. Oliver}
\affiliation{Research Laboratory of Electronics, Massachusetts Institute of Technology, 77 Massachusetts Avenue, Cambridge, Massachusetts 02139}
\affiliation{MIT Lincoln Laboratory, 244 Wood Street, Lexington, Massachusetts 02420, USA}
\begin{abstract}

Remarkable advancements in coherence and control fidelity have been achieved in recent years with cryogenic solid-state qubits. Nonetheless, thermalizing such devices to their milliKelvin environments has remained a long-standing fundamental and technical challenge.
In this context, we present a systematic study of the first-excited-state population in a 3D transmon superconducting qubit mounted in a dilution refrigerator with a variable temperature. Using a modified version of the protocol developed by Geerlings {\em et al.}~\cite{Yale_Temperature}, we observe the excited-state population to be consistent with a Maxwell-Boltzmann distribution, {\em i.e.}, a qubit in thermal equilibrium with the refrigerator, over the temperature range 35-150~mK. Below 35~mK, the excited-state population saturates at approximately 0.1\%. We verified this result using a flux qubit with ten-times stronger coupling to its readout resonator. We conclude that these qubits have effective temperature $T_{\mathrm{eff}} = 35$~mK. Assuming $T_{\mathrm{eff}}$ is due solely to hot quasiparticles, the inferred qubit lifetime is 108 $\mu$s and in plausible agreement with the measured 80 $\mu$s.
\end{abstract}
\date{\today}

\maketitle

Superconducting qubits are increasingly promising candidates to serve as the logic elements of a quantum information processor.
This assertion reflects, in part, several successes over the past decade 
addressing the fundamental operability of this qubit modality~\cite{DiVincenzo_arXiv,Devoret_Science_Review}.
A partial list includes a five-orders-of-magnitude increase in the coherence time $T_2$~\cite{Will_Review}, the active initialization of qubits in their ground state~\cite{Microwave_cooling, Yale_Temperature}, the demonstration of low-noise parametric amplifiers~\cite{PA_Colorado, PA_NEC, JPC_Yale1, PA_Berkeley, PA_Chalmers, PA_UCSB, TWPA_Berkeley} enabling high-fidelity readout~\cite{HFidel_Berkeley,HFidel_Delft,Lin2013a,HFidel_Yale}, and the implementation of a universal set of high-fidelity gates~\cite{Barends2014a}. 
In addition, prototypical quantum algorithms~\cite{Algorithms_Yale1, Algorithms_Yale2, Algorithms_UCSB} and simulations~\cite{Gustavsson2013a,Chen2014a} have been demonstrated with few-qubit systems, and the basic parity measurements underlying certain error detection protocols are now being realized with qubit stabilizers~\cite{Parity_DiCarlo1, Parity_DiCarlo2, Parity_IBM, Kelly15a_UCSB_Nature, Corcoles15a_arXiv, Riste15a_logical_qubit_NatComm} and photonic memories~\cite{Parity_Yale}.

Concomitant with these advances is an enhanced ability to improve our understanding of the technical and fundamental limitations of single qubits.
The 3D transmon~\cite{Paik2011a} has played an important role in this regard, because its relatively clean electromagnetic environment, predominantly low-loss qubit-mode volume, and resulting long coherence times make it a sensitive testbed for probing these limitations.

One such potential limitation is the degree to which a superconducting qubit is in equilibrium with its cryogenic environment.
Consider a typical superconducting qubit with a level splitting $E_{\mathrm{ge}} = h f_{\mathrm{ge}}$, with $f_{\mathrm{ge}} = 5 \; \mathrm{GHz}$, mounted in a dilution refrigerator at temperature $T=15$ mK, such that $E_{\mathrm{ge}} \gg k_{\mathrm{B}} T$.
Ideally, such a qubit in thermal equilibrium with the refrigerator will have a thermal population $\Pe \approx 10^{-5} \; \% $ of its first excited state according to Maxwell-Boltzmann statistics.
In practice, however, the empirical excited-state population reported for various superconducting qubits (featuring similar parameters $E_{\mathrm{ge}}$ and 
$T$) can be orders of magnitude higher, generally in the range of 1\%-13\% in steady state, corresponding to effective temperatures $T_{\mathrm{eff}} = 50-130$ mK~\cite{IBM_Temperature, Berkeley_Temperature, Delft_Temperature, Yale_Temperature}.

Thermalizing to milliKelvin temperatures has been a long-standing challenge for both normal and superconducting devices~\cite{Thermal1}.
A primary cause is thermal noise or blackbody radiation from higher temperature stages driving the device out of equilibrium, {\em e.g.,} via direct illumination or transferred via wires to the devices.
Several techniques have been identified to reduce these effects, including the use of microwave dissipative filters~\cite{Bladh03a} based on attenuation in meander lines~\cite{Vion1995a,Sueur06a}, fine-grain powders~\cite{Devoret1985a,Martinis1987a,Milliken07a,Lukashenko08a,Fukushima97a}, thin coaxial lines~\cite{Fukushima97a,Zorin1995a,Glattli1997a}, and lossy transmission lines~\cite{Courtois95a,Santavicca08a,Slichter09a}; differential mode operation~\cite{Vijay09a}; the importance of light-tight shielding practices~\cite{Hergenrother1995a}; and the introduction of low-reflectivity, infrared-absorbing (``black'') surface treatments~\cite{Persky99a}.
These techniques have been adapted to address qubit excited-state population 
by reducing stray or guided thermal photons~\cite{IBM_Temperature, UCSB_Wiring}.
Nonetheless, the problem is not fully eliminated and, moreover, the mechanism that generates residual excited-state population has yet to be clarified.

In this Letter, we report a systematic study of excited-state population in a 3D transmon qubit as measured in our system.
We developed a modified version of the protocol introduced by Geerlings \textit{et al.}~\cite{Yale_Temperature} to measure the excited-state population $\Pe$ as a function of bath temperature.
%
Our measurements are consistent with 
a qubit in thermal equilibrium with the dilution refrigerator over the temperature range 35-150 mK.
For temperatures below 35 mK, $\Pe$ saturates to a residual value of approximately 0.1\%, a factor 2.5 larger than the error of our measurement.
Ascribing this residual 
population entirely to non-equilibrium hot quasiparticles, the upper limit of quasiparticle density is estimated to be $2.2\times10^{-7}$ per Cooper pair.
The corresponding quasiparticle-induced decay time is calculated to be $T_1 = 108 \; \mu \mathrm{s}$, in reasonable agreement with the independently measured decay time $T_1=80 \; \mu \mathrm{s}$.
%
This suggests that both the residual excited-state population and relaxation times may be limited by
quasiparticles for this device.

The experiments were conducted in a Leiden cryogen-free dilution refrigerator (model CF-450) with a base temperature of 15 mK.
A temperature controller (model Lakeshore 370) is used to set the temperature with better than 0.1 mK stability at the thermometer.
A detailed schematic indicating the placement and types of attenuation and filters used in this
measurement is presented in the supplementary material~\cite{Supplementary}.

The sample is a $5\times5$ mm$^2$ sapphire chip comprising an aluminum, single-junction 3D transmon qubit~\cite{Paik2011a} with energy scales $E_J/E_C = 58$ and transition frequencies $f_{\rm ge}=4.97$ GHz and $f_{\rm ef}=4.70$ GHz.
The qubit is controlled using a circuit-QED approach through its strong dispersive coupling ($g / 2\pi = 160$ MHz) to an aluminum cavity with a TE101 mode frequency of 10.976 GHz (when loaded with a sapphire chip), an internal quality factor $Q_{\mathrm{i}} > 10^6$, and two ports with a net coupling $Q_{\mathrm{c}} = 10^5$.
The chip is mounted in the geometric center of the cavity using indium at the corners.
The sample in the present experiment exhibited coherence times: 
$T_1 = 80$ $\mu$s ($60-90$ $\mu$s),
$T_2^*=115$ $\mu$s ($90-115$ $\mu$s),
and $T_{2E}=154 \approx 2 T_1$ $\mu$s.
The observed range of $T_1$ and $T_2^*$ times over multiple cooldowns of this device are indicated parenthetically.
All the experiments presented in this paper are carried out with a standard dispersive readout method and without the use of a parametric amplifier.

In principle, when there is non-zero excited-state population $\Pe$ in the qubit, one should be able to observe an $e\rightarrow f$ transition peak in qubit spectroscopy.
In practice, however, it may be difficult to distinguish this transition experimentally from the background noise for small $\Pe$.
In a recent publication~\cite{Yale_Temperature}, Geerlings \textit{et al.} reported a method to measure small $\Pe$ levels ($\approx 1-10\%$, $T_{\mathrm{eff}}=60-100$ mK in their 3D transmon).
In their approach, $\Pe$ is determined by driving a Rabi oscillation between qubit states $\EE$ and $\FF$, hereafter called an ``{\em e-f} Rabi oscillation''.
In this work, we measured $\Pe$ using a modified protocol based on this method.

\begin{figure}[t]
    \includegraphics[width=7cm]{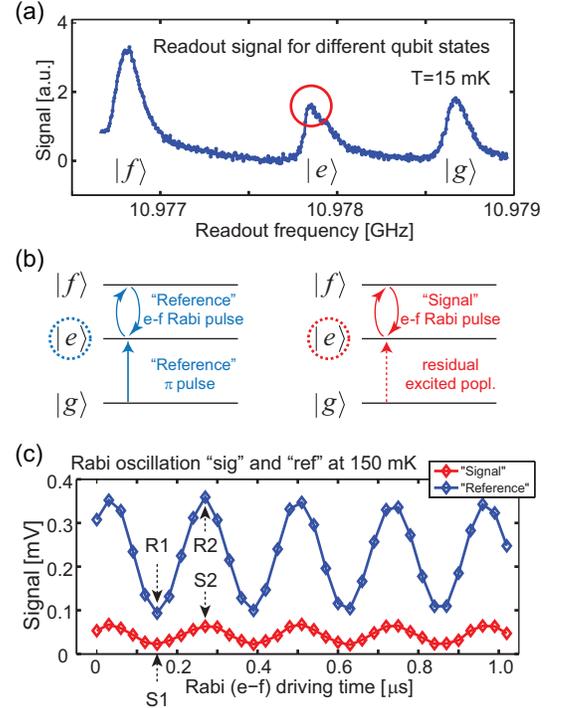}
    \caption{(a). The readout signal versus readout frequency. Three well-separated peaks are visible, corresponding to different qubit states. We read out state $|e\rangle$.
    (b) Modified experiment protocol using e-f Rabi driving. The excited state is populated by a $\pi^{g-e}$ pulse (left panel) or environmental excitation (right panel).
    (c). Observed e-f Rabi oscillations at 150 mK. The blue trace determines $\Aref$, while the red trace determines $\Asig$. When $\Asig$ is small, only two points on the blue trace (R1, R2) and the red trace (S1, S2) are measured. See text for details.}
    \label{Fig:readout_signal}
\end{figure}

In Fig.~\ref{Fig:readout_signal}a, the readout-signal amplitude as a function of readout-signal frequency indicates the dressed cavity frequency for states $|g \rangle$, $|e \rangle$, and $|f \rangle$.
For purposes of illustration, the qubit was prepared in state $|f \rangle$ using sequential $\pi^{\rm g \rightarrow e}$ and $\pi^{\rm e \rightarrow f}$ pulses, and then allowed to relax and partially populate states $|g \rangle$ and $|e \rangle$ before readout~\cite{Peterer15a_PRL}.

Whereas Geerlings {\em et al.} used a frequency corresponding to state $\GG$ for qubit readout, in our experiment, we use the readout frequency corresponding to state $\EE$ (red circle) to measure directly the {\em e-f} Rabi oscillation.
Reading out state $\EE$ simplifies the protocol by reducing the required number of $\pi^{\rm g \rightarrow e}$ pulses.
Moreover, since the readout tone for state $\EE$ is off-resonance with the cavity when the qubit is in state $\GG$, its predominant state in this experiment, and the cavity $Q$ is sufficiently high ($Q_c = 10^5$),
the cavity is only resonantly excited during readout in the rare cases that the qubit is in state $\EE$.

The modified measurement protocol is illustrated in Fig.~\ref{Fig:readout_signal}b.
We measure the {\em e-f} Rabi oscillation for two different conditions.
First, we apply a $\pi^{\rm g \rightarrow e}$ pulse to the qubit, swapping the populations of states $\GG$ and $\EE$ (left panel, Fig.~\ref{Fig:readout_signal}b).
We then apply an {\em e-f} driving pulse and read out state $|e\rangle$ as a function of the pulse duration.
The resulting Rabi oscillation is measured for 1 $\mu$s, containing more than 4 periods, and it appears sinusoidal due to the long Rabi decay time $T_R>100$ $\mu$s.
Note that the $\pi^{\rm g \rightarrow e}$ pulse swaps the populations of state $\EE$ and $\GG$.
Assuming the qubit population exists entirely within states $\GG$, $\EE$, and $\FF$, the oscillation amplitude is proportional to $\Pg - \Pf$, where $\Pg$ and $\Pf$ are the occupation probabilities of $\GG$ and $\FF$, respectively.
We denote this amplitude $\Aref$, the reference used when determining 
$\Pe$.

Second, we solely apply an {\em e-f} Rabi driving pulse without the $\pi^{\rm g \rightarrow e}$ pulse (right panel, Fig.~\ref{Fig:readout_signal}b).
In this case, the observed oscillation amplitude is proportional to $\Pe - \Pf$.
We denote the oscillation amplitude $\Asig$, the signal to be compared with the reference.
%

We are most interested in determining $\Pe$ in the low-temperature limit, {\em i.e.,} near the base temperature 15 mK.
At sufficiently low bath temperatures, {\em i.e.,} $T \ll E_{\rm ge} / k_{\mathrm{B}} \approx  E_{\rm ef} / k_{\mathrm{B}} \approx 235$ mK, we take $\Pf \rightarrow 0$ in our analytic treatment.
This assumption is reasonable, since one normally expects $\Pf \leq \Pe \leq \Pg$ in the absence of extraneous coherent excitation (we observe no evidence of such excitations).
Furthermore, simulated populations based on the Maxwell-Boltzmann distribution (see below) are consistent with this assumption for $T \leq 50$ mK.
It follows that $\Asig = A_0 \Pe$ and $\Aref = A_0 \Pg$, where $A_0$ is a factor converting the qubit state occupation probability to the readout voltage.
In this limit, $\Pe + \Pg = 1$ and $\Asig + \Aref = A_0$, such that the population of state $\EE$ is:
\begin{equation}
    \Pe^{\rm{exp}} \equiv \Asig / A_0 = {\Asig} / (\Asig + \Aref)
    \label{eq:Pe}
\end{equation}
in which $\Asig$ and $\Aref$ are determined experimentally. We emphasize that for $T\leq 50$ mK, $\Pe^{\rm{exp}}$ is a very good estimator for $\Pe$ in this device.

While measuring $\Aref$ is straightforward due to its large signal-to-noise ratio, the main technical challenge is to measure $\Asig$ precisely at the lowest temperatures.
When the population of state $\EE$ is in the range of 1-10\%~\cite{Yale_Temperature}, one can directly determine $\Asig$ by fitting the observed {\em e-f} Rabi trace to a sinusoidal function.
In our setup, a similarly discernable $\Pe$ level can be obtained by heating the sample to higher temperatures, where thermally excited population at state $\EE$ is significant.
In Fig.~\ref{Fig:readout_signal}c, the {\em e-f} Rabi trace with (blue points) and without (red points) the $\pi^{g \rightarrow e}$ swap pulse were both visible at an elevated bath temperature of 150 mK, enabling us to measure directly both $\Aref$ and $\Asig$.

In principle, provided one averages sufficiently, one can reduce the background noise and determine $\Asig$ using this trace-fitting method.
However, assuming that each experiment is independent, the background fluctuations decrease only as the square-root of the number of trials averaged.
Improving the resolution from 1\% to 0.1\% would require a factor $100 \times$ more trials and, thus, a factor $100 \times$ in time.
As a result, for $\Pe \ll 1\%$, it is practically prohibitive to measure the entire trace in Fig.~\ref{Fig:readout_signal}c ({\em i.e.,} 35 points, each requiring approximately $10^7$ averages given our set-up).

We therefore further modified the experimental protocol to increase data acquisition efficiency.
Since we use the same {\em e-f} Rabi driving power to measure both the signal and the reference traces, we expect and confirmed the frequency and phase of these traces to be the same.
We can therefore obtain amplitudes $\Asig$ and $\Aref$ by measuring two points each: the maximum S1 and minimum S2 amplitudes for the signal trace and, similarly, R1 and R2 for the reference trace~\cite{Supplementary}.
Compared with measuring the full trace, this "two-point" method greatly reduces the acquisition time.

%
\begin{figure}[t]
    \includegraphics[width=7cm]{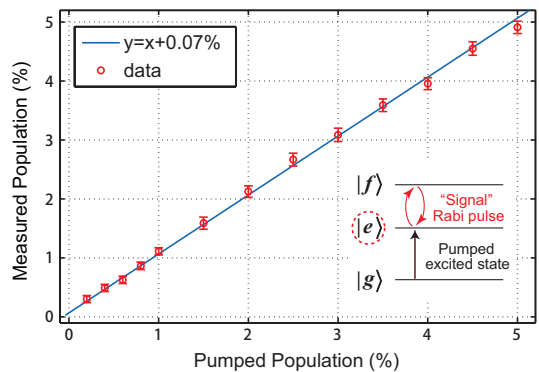}
    \caption{Calibration measurement of excited state population (percent), pumped using a small fraction of a $\pi^{\rm g-e}$ pulse. The data can be fit to a linear $y=x+b$ function, where $b=0.067 \%$, with 95\% confidence bounds of $(0.025 \%, 0.011 \%)$. }
    \label{Fig:validation}
\end{figure}

We designed a calibration experiment to validate the protocol.
We first applied a small fraction of a $\pi^{g \rightarrow e}$ pulse to the qubit, which pumps $k$\% of the ground-state population $\Pg$ to state $|e\rangle$, and simultaneously brings $k$\% of any residual excited-state population $\Pe$ to ground state.
The pumped excited-state population $\Pe^p$ is
\begin{equation}
    \Pe^p = k\Pg + (1-k)\Pe 
    \label{eq:pumping}
\end{equation}
in which 
$\Pe$ is the initial excited-state population.
We then drove an {\em e-f} Rabi oscillation and measured the oscillation amplitude $\Asig$.
The measured $\Pe^p$ should depend linearly on $k$, and its intercept at $k=0$ (no pumping pulse) is $\Pe$ at base temperature ({\em i.e.}, assuming $\Pf = 0$).

We scanned $k$ over the range $0.2\% - 5.0\%$ and measured $\Pe^p$ at the base temperature $T=15$ mK (see Fig.~\ref{Fig:validation}).
The data 
fit well to a linear function, validating the protocol,
%
and yield an intercept $\Pe=0.067\%$, with 95\% confidence bounds of 
$(0.025\%, 0.011\%)$
This value can in fact be regarded as one estimate for the residual excited-state population at the bath temperature of 15 mK.

\begin{figure}[t]
    \includegraphics[width=8cm]{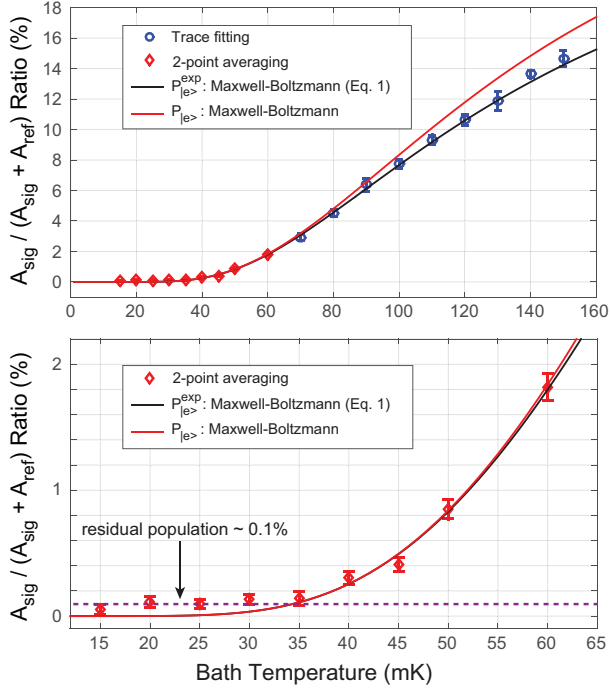}
    \caption{
    (a) $\Pe^{\rm{exp}}$ ratio (Eq.~\ref{eq:Pe}) versus temperature, 15-150 mK. Experimental data are obtained through fitting a 1-$\mu$s Rabi trace (blue points) or the two-point method (red points). Solid lines are calculated $\Pe^{\rm{exp}}$ (blue line) and $\Pe$ (red line) based on the Maxwell-Boltzmann distribution for the lowest four energy levels (see text).
    (b) Zoom: $\Pe^{\rm{exp}}$ ratio versus temperature, 15-60 mK. In this limit, $\Pe^{\rm{exp}}$ is a good estimator for $\Pe$   The data saturate to 0.1\% at lower temperatures (purple dashed line) with an inferred effective temperature of 35 mK.}
    \label{Fig:temp_dependence}
\end{figure}

When the bath temperature is raised, one expects that the excited-state population of the qubit will increase (see Fig.\ref{Fig:readout_signal}c)
In thermal equilibrium with the refrigerator at temperature $T$, the qubit-state population of states  $|i\rangle$ at energies $E_i$ follow a Maxwell-Boltzmann distribution,
\begin{equation}
    P_{|i\rangle}= \frac{1}{Z}g_i\exp(-E_i/k_BT).
    \label{eq:Boltzmann}
\end{equation}
Here, $Z = \sum_j g_j \exp(-E_j/k_BT)$ is the partition function, $g_i$ is the degeneracy of each energy level $E_i$, and $k_B$ is the Boltzmann constant.
In our analysis, we define $E_{|g\rangle} \equiv 0$, $g_i = 1$, and consider the lowest-four energy levels in the transmon (a sufficient number for the temperature range considered here)~\cite{Peterer15a_PRL}.
Using Eq.~(\ref{eq:Boltzmann}), we calculate the equilibrium population $\Pe$ and the ratio $\Pe^{\rm{exp}}$ (see Eq.~\ref{eq:Pe}) versus temperature, and plot them in Figs.~\ref{Fig:temp_dependence}a and~\ref{Fig:temp_dependence}b.
The equilibrium traces $\Pe^{\rm{exp}}$ and $\Pe$ are indistinguishable for $T \leq 50 $ mK.
At higher temperatures the assumption $\Pf=0$ is no longer valid, and the traces differ by as much as 2\% at 160 mK.

Excited-state population measurements were performed as a function of temperature over the range $T = 15-150$ mK.
For each set point, after the temperature sensor (fixed on the cold finger near the device) reading is stable to within 0.1 mK, we wait an additional 2 hours before acquiring data to ensure the qubit has reached its steady-state population distribution.
In Fig.~\ref{Fig:temp_dependence}a the experimental $\Pe^{\rm{exp}}$ generally matches the simulation of Eq.~\ref{eq:Pe} assuming Maxwell-Boltzmann populations (black trace) over the range 35-150 mK, consistent with the qubit being in thermal equilibrium with the cryostat.
In the range 35-60 mK, $\Pe^{\rm{exp}}$ also matches the Maxwell-Boltzmann estimate for $\Pe$ (red trace).
Below 35 mK (Fig.~\ref{Fig:temp_dependence}b), the experimental $\Pe^{\rm{exp}}$ deviates from thermal equilibrium, saturating at approximately $\Pe^{\rm{exp}} = \Pe = 0.1\%$ (purple dashed line).
That is, $\Pe^{\rm{exp}} \leq 0.1\% + \Pf$ and becomes $\Pe^{\rm{exp}} = 0.1\%$ with the reasonable assumption $\Pf=0$ (see Eq.~\ref{eq:Pe}).
This saturation level is consistent with the 0.067\% estimate obtained during the calibration experiment (Fig~\ref{Fig:validation}).
Although $\Pe= 0.1\%$ is an order of magnitude lower than other reports in the literature, it remains about four orders of magnitude higher than the expected equilibrium value ($\sim 10^{-5} \%$) at 15 mK.
We note that we used a level of averaging sufficient to achieve small (0.04\%) error bars on the population of $0.1 \%$.
In addition to more averaging, using a low-noise parametric amplifier would further improve the signal-to-noise ratio and allow for single-shot readout with higher resilience to low-frequency noise~\cite{Supplementary}.

We define an effective temperature $\Teff$ as the temperature that would have generated the observed $\Pe$ in an otherwise identical equilibrium qubit, according to Eq.~\ref{eq:Boltzmann}.
In our qubit, the cross-over from thermal equilibrium to saturation at $\Pe=0.1\%$ occurs at $\Teff = 35$ mK.

A potential mechanism for the observed non-equilibrium qubit temperature is the presence of ``hot'' non-equilibrium quasiparticles ({\em i.e.,} those with energy higher than $\Delta + E_{\rm ge}$, where $\Delta$ is the superconducting energy gap)~\cite{UCSB_Hot_QP}.
Stray thermal photons entering the cavity from higher-temperature stages of the refrigerator may in principle generate new quasiparticles or heat existing ones depending on the photon energy.
Such ``hot'' quasiparticles, in turn, lose energy $E_{\rm ge}$ to the qubit and drive it out of thermal equilibrium to a degree determined by the non-equilibrium quasiparticle density. 
Following Wenner {\em et al.,} the quasiparticle-induced excited-state population can be written as~\cite{UCSB_Hot_QP}
\begin{equation}
    \Pe^{\rm qp} \simeq 2.17 (n_{\rm qp}/n_{\rm cp}) (\Delta/E_{\rm ge})^{3.65} \label{eq:hot_qp}
\end{equation}
in which $n_{\rm cp}$ is the Cooper-pair density and $n_{\rm qp}$ is the density of all quasiparticles.
Taking the observed excited-state population $\Pe^{\rm qp} = 0.1\%$ to be solely induced by quasiparticles, the upper limit for the  quasiparticle density is $(n_{\rm qp}/n_{\rm cp}) = 2.2\times10^{-7}$ per Cooper pair.

Within these assumptions, the quasiparticle-induced decay rate for a transmon qubit is~\cite{UCSB_Hot_QP,Catelani2011a}
\begin{equation}
    \Gamma_{\rm qp} \simeq \frac{\sqrt{2}}{R_{\mathrm{N}} C} \left( \frac{\Delta}{E_{\rm ge}} \right)^{3/2} \frac{n_{\rm qp}}{n_{\rm cp}}
\end{equation}
in which $R_{\mathrm{N}}$ is the normal-state resistance of the Josephson junction, and $C$ is the qubit capacitance.
Taking $\Delta=170 $ $\mu$eV, $R_{\mathrm{N}}=9.5$ k$\Omega$ and $C=80$ fF~\cite{Peterer15a_PRL, Supplementary}, we have $\Gamma_{\rm qp}=9.30$ kHz, corresponding to a relaxation time $T_1^{\rm qp}=108$ $\mu$s, which is only about 35\% larger than the measured time $T_1=80$ $\mu$s for this sample.

We have measured similar effective temperatures $T_{\mathrm{eff}} = 30-45$ mK for several superconducting qubit modalities (flux qubit, capacitively shunted flux qubit, 2D transmons) measured in our lab in both a dry (Leiden CF-450) and a wet (Oxford Kelvinox 400) refrigerator with similar wiring and filtering configurations~\cite{Supplementary}.
In particular, we observed $T = 35 \pm 4$ mK for a capacitively shunted flux qubit with similar qubit parameters, including $f_{\mathrm{ge}} = 4.7$ GHz, $f_\mathrm{resonator}=8.3$ GHz, $Q_{\mathrm{c}}=5000$ and $g/2 \pi = 100$ MHz.
This is notable, because this device was read out dispersively using a cavity with $10 \times$ lower $Q_{\mathrm{c}}$, that is, with a much stronger coupling to the coaxial cables in our refrigerator than the 3D transmon.
%

To summarize, we have studied the first-excited-state population of a 3D transmon qubit over the temperature range $T=15-150$ mK.
The excited-state population matches Maxwell-Boltzmann statistics over the range $T = 35-150$ mK, consistent with a qubit in thermal equilibrium with the refrigerator.
For temperatures below 35 mK, the excited-state population saturates to a small value of approximately 0.1\%.
Assuming the residual population is solely caused by non-equilibrium ``hot'' quasiparticles, the calculated and measured relaxation times are plausibly consistent for this device.
We have observed similarly low effective temperature in multiple devices and configurations, including a readout resonator with $10 \times$ larger coupling Q.
While we present our full filtering and attenuation schematic in the supplementary material~\cite{Supplementary}, we did not need to change any particular aspect of our measurement system to achieve these effective temperatures, and so there is no particular ``reason'' beyond careful cryogenic engineering that we can identify for their relatively low values.

We thank George Fitch and Terry Weir for technical assistance and A.J. Kerman for useful discussions. This research was funded in part by the Assistant Secretary of Defense for Research \& Engineering under Air Force Contract FA8721-05-C-0002, by the U.S. Army Research Office (W911NF-12-1-0036), and by the National Science Foundation (PHY-1005373).

\nocite{Yan2014a}
\nocite{Ambegaokar63a}
\nocite{Ambegaokar63b}

\bibliographystyle{apsrev4-1}
\bibliography{population_V4}

\begin{thebibliography}{58}%
\makeatletter
\providecommand \@ifxundefined [1]{%
 \@ifx{#1\undefined}
}%
\providecommand \@ifnum [1]{%
 \ifnum #1\expandafter \@firstoftwo
 \else \expandafter \@secondoftwo
 \fi
}%
\providecommand \@ifx [1]{%
 \ifx #1\expandafter \@firstoftwo
 \else \expandafter \@secondoftwo
 \fi
}%
\providecommand \natexlab [1]{#1}%
\providecommand \enquote  [1]{``#1''}%
\providecommand \bibnamefont  [1]{#1}%
\providecommand \bibfnamefont [1]{#1}%
\providecommand \citenamefont [1]{#1}%
\providecommand \href@noop [0]{\@secondoftwo}%
\providecommand \href [0]{\begingroup \@sanitize@url \@href}%
\providecommand \@href[1]{\@@startlink{#1}\@@href}%
\providecommand \@@href[1]{\endgroup#1\@@endlink}%
\providecommand \@sanitize@url [0]{\catcode `\\12\catcode `\$12\catcode
  `\&12\catcode `\#12\catcode `\^12\catcode `\_12\catcode `\%12\relax}%
\providecommand \@@startlink[1]{}%
\providecommand \@@endlink[0]{}%
\providecommand \url  [0]{\begingroup\@sanitize@url \@url }%
\providecommand \@url [1]{\endgroup\@href {#1}{\urlprefix }}%
\providecommand \urlprefix  [0]{URL }%
\providecommand \Eprint [0]{\href }%
\providecommand \doibase [0]{http://dx.doi.org/}%
\providecommand \selectlanguage [0]{\@gobble}%
\providecommand \bibinfo  [0]{\@secondoftwo}%
\providecommand \bibfield  [0]{\@secondoftwo}%
\providecommand \translation [1]{[#1]}%
\providecommand \BibitemOpen [0]{}%
\providecommand \bibitemStop [0]{}%
\providecommand \bibitemNoStop [0]{.\EOS\space}%
\providecommand \EOS [0]{\spacefactor3000\relax}%
\providecommand \BibitemShut  [1]{\csname bibitem#1\endcsname}%
\let\auto@bib@innerbib\@empty
\bibitem [{\citenamefont {Geerlings}\ \emph {et~al.}(2013)\citenamefont
  {Geerlings}, \citenamefont {Leghtas}, \citenamefont {Pop}, \citenamefont
  {Shankar}, \citenamefont {Frunzio}, \citenamefont {Schoelkopf}, \citenamefont
  {Mirrahimi},\ and\ \citenamefont {Devoret}}]{Yale_Temperature}%
  \BibitemOpen
  \bibfield  {author} {\bibinfo {author} {\bibfnamefont {K.}~\bibnamefont
  {Geerlings}}, \bibinfo {author} {\bibfnamefont {Z.}~\bibnamefont {Leghtas}},
  \bibinfo {author} {\bibfnamefont {I.~M.}\ \bibnamefont {Pop}}, \bibinfo
  {author} {\bibfnamefont {S.}~\bibnamefont {Shankar}}, \bibinfo {author}
  {\bibfnamefont {L.}~\bibnamefont {Frunzio}}, \bibinfo {author} {\bibfnamefont
  {R.~J.}\ \bibnamefont {Schoelkopf}}, \bibinfo {author} {\bibfnamefont
  {M.}~\bibnamefont {Mirrahimi}}, \ and\ \bibinfo {author} {\bibfnamefont
  {M.~H.}\ \bibnamefont {Devoret}},\ }\href@noop {} {\bibfield  {journal}
  {\bibinfo  {journal} {Phys. Rev. Lett.}\ }\textbf {\bibinfo {volume} {110}},\
  \bibinfo {pages} {120501} (\bibinfo {year} {2013})}\BibitemShut {NoStop}%
\bibitem [{\citenamefont {DiVincenzo}(2000)}]{DiVincenzo_arXiv}%
  \BibitemOpen
  \bibfield  {author} {\bibinfo {author} {\bibfnamefont {D.~P.}\ \bibnamefont
  {DiVincenzo}},\ }\href@noop {} {\bibfield  {journal} {\bibinfo  {journal}
  {arXiv:quant-ph/0002077v3}\ } (\bibinfo {year} {2000})}\BibitemShut {NoStop}%
\bibitem [{\citenamefont {Devoret}\ and\ \citenamefont
  {Schoelkopf}(2013)}]{Devoret_Science_Review}%
  \BibitemOpen
  \bibfield  {author} {\bibinfo {author} {\bibfnamefont {M.~H.}\ \bibnamefont
  {Devoret}}\ and\ \bibinfo {author} {\bibfnamefont {R.~J.}\ \bibnamefont
  {Schoelkopf}},\ }\href@noop {} {\bibfield  {journal} {\bibinfo  {journal}
  {Science}\ }\textbf {\bibinfo {volume} {339}},\ \bibinfo {pages} {1169}
  (\bibinfo {year} {2013})}\BibitemShut {NoStop}%
\bibitem [{\citenamefont {Oliver}\ and\ \citenamefont
  {Welander}(2013)}]{Will_Review}%
  \BibitemOpen
  \bibfield  {author} {\bibinfo {author} {\bibfnamefont {W.~D.}\ \bibnamefont
  {Oliver}}\ and\ \bibinfo {author} {\bibfnamefont {P.~B.}\ \bibnamefont
  {Welander}},\ }\href@noop {} {\bibfield  {journal} {\bibinfo  {journal} {MRS
  Bulletin}\ }\textbf {\bibinfo {volume} {38}},\ \bibinfo {pages} {816}
  (\bibinfo {year} {2013})}\BibitemShut {NoStop}%
\bibitem [{\citenamefont {Valenzuela}\ \emph {et~al.}(2006)\citenamefont
  {Valenzuela}, \citenamefont {Oliver}, \citenamefont {Berns}, \citenamefont
  {Berggren}, \citenamefont {Levitov},\ and\ \citenamefont
  {Orlando}}]{Microwave_cooling}%
  \BibitemOpen
  \bibfield  {author} {\bibinfo {author} {\bibfnamefont {S.~O.}\ \bibnamefont
  {Valenzuela}}, \bibinfo {author} {\bibfnamefont {W.~D.}\ \bibnamefont
  {Oliver}}, \bibinfo {author} {\bibfnamefont {D.~M.}\ \bibnamefont {Berns}},
  \bibinfo {author} {\bibfnamefont {K.~K.}\ \bibnamefont {Berggren}}, \bibinfo
  {author} {\bibfnamefont {L.~S.}\ \bibnamefont {Levitov}}, \ and\ \bibinfo
  {author} {\bibfnamefont {T.~P.}\ \bibnamefont {Orlando}},\ }\href@noop {}
  {\bibfield  {journal} {\bibinfo  {journal} {Science}\ }\textbf {\bibinfo
  {volume} {314}},\ \bibinfo {pages} {1589} (\bibinfo {year}
  {2006})}\BibitemShut {NoStop}%
\bibitem [{\citenamefont {Castellanos-Beltran}\ and\ \citenamefont
  {Lehnert}(2007)}]{PA_Colorado}%
  \BibitemOpen
  \bibfield  {author} {\bibinfo {author} {\bibfnamefont {M.~A.}\ \bibnamefont
  {Castellanos-Beltran}}\ and\ \bibinfo {author} {\bibfnamefont {K.~W.}\
  \bibnamefont {Lehnert}},\ }\href@noop {} {\bibfield  {journal} {\bibinfo
  {journal} {Appl. Phys. Lett.}\ }\textbf {\bibinfo {volume} {91}},\ \bibinfo
  {pages} {083509} (\bibinfo {year} {2007})}\BibitemShut {NoStop}%
\bibitem [{\citenamefont {Yamamoto}\ \emph {et~al.}(2008)\citenamefont
  {Yamamoto}, \citenamefont {Inomata}, \citenamefont {Watanabe}, \citenamefont
  {Matsuba}, \citenamefont {Miyazaki}, \citenamefont {Oliver}, \citenamefont
  {Nakamura},\ and\ \citenamefont {Tsai}}]{PA_NEC}%
  \BibitemOpen
  \bibfield  {author} {\bibinfo {author} {\bibfnamefont {T.}~\bibnamefont
  {Yamamoto}}, \bibinfo {author} {\bibfnamefont {K.}~\bibnamefont {Inomata}},
  \bibinfo {author} {\bibfnamefont {M.}~\bibnamefont {Watanabe}}, \bibinfo
  {author} {\bibfnamefont {K.}~\bibnamefont {Matsuba}}, \bibinfo {author}
  {\bibfnamefont {T.}~\bibnamefont {Miyazaki}}, \bibinfo {author}
  {\bibfnamefont {W.~D.}\ \bibnamefont {Oliver}}, \bibinfo {author}
  {\bibfnamefont {Y.}~\bibnamefont {Nakamura}}, \ and\ \bibinfo {author}
  {\bibfnamefont {J.}~\bibnamefont {Tsai}},\ }\href@noop {} {\bibfield
  {journal} {\bibinfo  {journal} {Appl. Phys. Lett.}\ }\textbf {\bibinfo
  {volume} {93}},\ \bibinfo {pages} {042510} (\bibinfo {year}
  {2008})}\BibitemShut {NoStop}%
\bibitem [{\citenamefont {Bergeal}\ \emph {et~al.}(2010)\citenamefont
  {Bergeal}, \citenamefont {Schackert}, \citenamefont {Metcalfe}, \citenamefont
  {Vijay}, \citenamefont {Manucharyan}, \citenamefont {Frunzio}, \citenamefont
  {Prober}, \citenamefont {Schoelkopf}, \citenamefont {Girvin},\ and\
  \citenamefont {Devoret}}]{JPC_Yale1}%
  \BibitemOpen
  \bibfield  {author} {\bibinfo {author} {\bibfnamefont {N.}~\bibnamefont
  {Bergeal}}, \bibinfo {author} {\bibfnamefont {F.}~\bibnamefont {Schackert}},
  \bibinfo {author} {\bibfnamefont {M.}~\bibnamefont {Metcalfe}}, \bibinfo
  {author} {\bibfnamefont {R.}~\bibnamefont {Vijay}}, \bibinfo {author}
  {\bibfnamefont {V.~E.}\ \bibnamefont {Manucharyan}}, \bibinfo {author}
  {\bibfnamefont {L.}~\bibnamefont {Frunzio}}, \bibinfo {author} {\bibfnamefont
  {D.~E.}\ \bibnamefont {Prober}}, \bibinfo {author} {\bibfnamefont {R.~J.}\
  \bibnamefont {Schoelkopf}}, \bibinfo {author} {\bibfnamefont {S.~M.}\
  \bibnamefont {Girvin}}, \ and\ \bibinfo {author} {\bibfnamefont {M.~H.}\
  \bibnamefont {Devoret}},\ }\href@noop {} {\bibfield  {journal} {\bibinfo
  {journal} {Nature}\ }\textbf {\bibinfo {volume} {465}},\ \bibinfo {pages}
  {64} (\bibinfo {year} {2010})}\BibitemShut {NoStop}%
\bibitem [{\citenamefont {Hatridge}\ \emph {et~al.}(2011)\citenamefont
  {Hatridge}, \citenamefont {Vijay}, \citenamefont {Slichter}, \citenamefont
  {Clarke},\ and\ \citenamefont {Siddiqi}}]{PA_Berkeley}%
  \BibitemOpen
  \bibfield  {author} {\bibinfo {author} {\bibfnamefont {M.}~\bibnamefont
  {Hatridge}}, \bibinfo {author} {\bibfnamefont {R.}~\bibnamefont {Vijay}},
  \bibinfo {author} {\bibfnamefont {D.~H.}\ \bibnamefont {Slichter}}, \bibinfo
  {author} {\bibfnamefont {J.}~\bibnamefont {Clarke}}, \ and\ \bibinfo {author}
  {\bibfnamefont {I.}~\bibnamefont {Siddiqi}},\ }\href@noop {} {\bibfield
  {journal} {\bibinfo  {journal} {Physical Review B}\ }\textbf {\bibinfo
  {volume} {83}},\ \bibinfo {pages} {134501} (\bibinfo {year}
  {2011})}\BibitemShut {NoStop}%
\bibitem [{\citenamefont {Sundqvist}\ \emph {et~al.}(2013)\citenamefont
  {Sundqvist}, \citenamefont {Kinta{\c{s}}}, \citenamefont {Simoen},
  \citenamefont {Krantz}, \citenamefont {Sandberg}, \citenamefont {Wilson},\
  and\ \citenamefont {Delsing}}]{PA_Chalmers}%
  \BibitemOpen
  \bibfield  {author} {\bibinfo {author} {\bibfnamefont {K.~M.}\ \bibnamefont
  {Sundqvist}}, \bibinfo {author} {\bibfnamefont {S.}~\bibnamefont
  {Kinta{\c{s}}}}, \bibinfo {author} {\bibfnamefont {M.}~\bibnamefont
  {Simoen}}, \bibinfo {author} {\bibfnamefont {P.}~\bibnamefont {Krantz}},
  \bibinfo {author} {\bibfnamefont {M.}~\bibnamefont {Sandberg}}, \bibinfo
  {author} {\bibfnamefont {C.~M.}\ \bibnamefont {Wilson}}, \ and\ \bibinfo
  {author} {\bibfnamefont {P.}~\bibnamefont {Delsing}},\ }\href@noop {}
  {\bibfield  {journal} {\bibinfo  {journal} {Appl. Phys. Lett.}\ }\textbf
  {\bibinfo {volume} {103}},\ \bibinfo {pages} {102603} (\bibinfo {year}
  {2013})}\BibitemShut {NoStop}%
\bibitem [{\citenamefont {Mutus}\ \emph {et~al.}(2013)\citenamefont {Mutus},
  \citenamefont {White}, \citenamefont {Jeffrey}, \citenamefont {Sank},
  \citenamefont {Barends}, \citenamefont {Bochmann}, \citenamefont {Chen},
  \citenamefont {Chen}, \citenamefont {Chiaro}, \citenamefont {Dunsworth} \emph
  {et~al.}}]{PA_UCSB}%
  \BibitemOpen
  \bibfield  {author} {\bibinfo {author} {\bibfnamefont {J.~Y.}\ \bibnamefont
  {Mutus}}, \bibinfo {author} {\bibfnamefont {T.~C.}\ \bibnamefont {White}},
  \bibinfo {author} {\bibfnamefont {E.}~\bibnamefont {Jeffrey}}, \bibinfo
  {author} {\bibfnamefont {D.}~\bibnamefont {Sank}}, \bibinfo {author}
  {\bibfnamefont {R.}~\bibnamefont {Barends}}, \bibinfo {author} {\bibfnamefont
  {J.}~\bibnamefont {Bochmann}}, \bibinfo {author} {\bibfnamefont
  {Y.}~\bibnamefont {Chen}}, \bibinfo {author} {\bibfnamefont {Z.}~\bibnamefont
  {Chen}}, \bibinfo {author} {\bibfnamefont {B.}~\bibnamefont {Chiaro}},
  \bibinfo {author} {\bibfnamefont {A.}~\bibnamefont {Dunsworth}},  \emph
  {et~al.},\ }\href@noop {} {\bibfield  {journal} {\bibinfo  {journal} {Appl.
  Phys. Lett.}\ }\textbf {\bibinfo {volume} {103}},\ \bibinfo {pages} {122602}
  (\bibinfo {year} {2013})}\BibitemShut {NoStop}%
\bibitem [{\citenamefont {O'Brien}\ \emph {et~al.}(2014)\citenamefont
  {O'Brien}, \citenamefont {Macklin}, \citenamefont {Siddiqi},\ and\
  \citenamefont {Zhang}}]{TWPA_Berkeley}%
  \BibitemOpen
  \bibfield  {author} {\bibinfo {author} {\bibfnamefont {K.}~\bibnamefont
  {O'Brien}}, \bibinfo {author} {\bibfnamefont {C.}~\bibnamefont {Macklin}},
  \bibinfo {author} {\bibfnamefont {I.}~\bibnamefont {Siddiqi}}, \ and\
  \bibinfo {author} {\bibfnamefont {X.}~\bibnamefont {Zhang}},\ }\href@noop {}
  {\bibfield  {journal} {\bibinfo  {journal} {Phys. Rev. Lett.}\ }\textbf
  {\bibinfo {volume} {113}},\ \bibinfo {pages} {157001} (\bibinfo {year}
  {2014})}\BibitemShut {NoStop}%
\bibitem [{\citenamefont {Vijay}\ \emph {et~al.}(2011)\citenamefont {Vijay},
  \citenamefont {Slichter},\ and\ \citenamefont {Siddiqi}}]{HFidel_Berkeley}%
  \BibitemOpen
  \bibfield  {author} {\bibinfo {author} {\bibfnamefont {R.}~\bibnamefont
  {Vijay}}, \bibinfo {author} {\bibfnamefont {D.~H.}\ \bibnamefont {Slichter}},
  \ and\ \bibinfo {author} {\bibfnamefont {I.}~\bibnamefont {Siddiqi}},\
  }\href@noop {} {\bibfield  {journal} {\bibinfo  {journal} {Phys. Rev. Lett.}\
  }\textbf {\bibinfo {volume} {106}},\ \bibinfo {pages} {110502} (\bibinfo
  {year} {2011})}\BibitemShut {NoStop}%
\bibitem [{\citenamefont {De~Lange}\ \emph {et~al.}(2014)\citenamefont
  {De~Lange}, \citenamefont {Rist{\`e}}, \citenamefont {Tiggelman},
  \citenamefont {Eichler}, \citenamefont {Tornberg}, \citenamefont {Johansson},
  \citenamefont {Wallraff}, \citenamefont {Schouten},\ and\ \citenamefont
  {DiCarlo}}]{HFidel_Delft}%
  \BibitemOpen
  \bibfield  {author} {\bibinfo {author} {\bibfnamefont {G.}~\bibnamefont
  {De~Lange}}, \bibinfo {author} {\bibfnamefont {D.}~\bibnamefont {Rist{\`e}}},
  \bibinfo {author} {\bibfnamefont {M.}~\bibnamefont {Tiggelman}}, \bibinfo
  {author} {\bibfnamefont {C.}~\bibnamefont {Eichler}}, \bibinfo {author}
  {\bibfnamefont {L.}~\bibnamefont {Tornberg}}, \bibinfo {author}
  {\bibfnamefont {G.}~\bibnamefont {Johansson}}, \bibinfo {author}
  {\bibfnamefont {A.}~\bibnamefont {Wallraff}}, \bibinfo {author}
  {\bibfnamefont {R.}~\bibnamefont {Schouten}}, \ and\ \bibinfo {author}
  {\bibfnamefont {L.}~\bibnamefont {DiCarlo}},\ }\href@noop {} {\bibfield
  {journal} {\bibinfo  {journal} {Phys. Rev. Lett.}\ }\textbf {\bibinfo
  {volume} {112}},\ \bibinfo {pages} {080501} (\bibinfo {year}
  {2014})}\BibitemShut {NoStop}%
\bibitem [{\citenamefont {Lin}\ \emph {et~al.}(2013)\citenamefont {Lin},
  \citenamefont {Inomata}, \citenamefont {Oliver}, \citenamefont {Koshino},
  \citenamefont {Nakamura}, \citenamefont {Tsai},\ and\ \citenamefont
  {T}}]{Lin2013a}%
  \BibitemOpen
  \bibfield  {author} {\bibinfo {author} {\bibfnamefont {Z.}~\bibnamefont
  {Lin}}, \bibinfo {author} {\bibfnamefont {K.}~\bibnamefont {Inomata}},
  \bibinfo {author} {\bibfnamefont {W.~D.}\ \bibnamefont {Oliver}}, \bibinfo
  {author} {\bibfnamefont {K.}~\bibnamefont {Koshino}}, \bibinfo {author}
  {\bibfnamefont {Y.}~\bibnamefont {Nakamura}}, \bibinfo {author}
  {\bibfnamefont {J.~S.}\ \bibnamefont {Tsai}}, \ and\ \bibinfo {author}
  {\bibfnamefont {Y.}~\bibnamefont {T}},\ }\href@noop {} {\bibfield  {journal}
  {\bibinfo  {journal} {Appl. Phys. Lett.}\ }\textbf {\bibinfo {volume}
  {103}},\ \bibinfo {pages} {132602} (\bibinfo {year} {2013})}\BibitemShut
  {NoStop}%
\bibitem [{\citenamefont {Abdo}\ \emph {et~al.}(2014)\citenamefont {Abdo},
  \citenamefont {Sliwa}, \citenamefont {Shankar}, \citenamefont {Hatridge},
  \citenamefont {Frunzio}, \citenamefont {Schoelkopf},\ and\ \citenamefont
  {Devoret}}]{HFidel_Yale}%
  \BibitemOpen
  \bibfield  {author} {\bibinfo {author} {\bibfnamefont {B.}~\bibnamefont
  {Abdo}}, \bibinfo {author} {\bibfnamefont {K.}~\bibnamefont {Sliwa}},
  \bibinfo {author} {\bibfnamefont {S.}~\bibnamefont {Shankar}}, \bibinfo
  {author} {\bibfnamefont {M.}~\bibnamefont {Hatridge}}, \bibinfo {author}
  {\bibfnamefont {L.}~\bibnamefont {Frunzio}}, \bibinfo {author} {\bibfnamefont
  {R.}~\bibnamefont {Schoelkopf}}, \ and\ \bibinfo {author} {\bibfnamefont
  {M.}~\bibnamefont {Devoret}},\ }\href@noop {} {\bibfield  {journal} {\bibinfo
   {journal} {Phys. Rev. Lett.}\ }\textbf {\bibinfo {volume} {112}},\ \bibinfo
  {pages} {167701} (\bibinfo {year} {2014})}\BibitemShut {NoStop}%
\bibitem [{\citenamefont {Barends}\ \emph {et~al.}(2014)\citenamefont
  {Barends}, \citenamefont {Kelly}, \citenamefont {Megrant}, \citenamefont
  {Veitia}, \citenamefont {Sank}, \citenamefont {Jeffrey}, \citenamefont
  {White}, \citenamefont {Mutus}, \citenamefont {Fowler}, \citenamefont
  {Campbell} \emph {et~al.}}]{Barends2014a}%
  \BibitemOpen
  \bibfield  {author} {\bibinfo {author} {\bibfnamefont {R.}~\bibnamefont
  {Barends}}, \bibinfo {author} {\bibfnamefont {J.}~\bibnamefont {Kelly}},
  \bibinfo {author} {\bibfnamefont {A.}~\bibnamefont {Megrant}}, \bibinfo
  {author} {\bibfnamefont {A.}~\bibnamefont {Veitia}}, \bibinfo {author}
  {\bibfnamefont {D.}~\bibnamefont {Sank}}, \bibinfo {author} {\bibfnamefont
  {E.}~\bibnamefont {Jeffrey}}, \bibinfo {author} {\bibfnamefont {T.~C.}\
  \bibnamefont {White}}, \bibinfo {author} {\bibfnamefont {J.}~\bibnamefont
  {Mutus}}, \bibinfo {author} {\bibfnamefont {A.~G.}\ \bibnamefont {Fowler}},
  \bibinfo {author} {\bibfnamefont {B.}~\bibnamefont {Campbell}},  \emph
  {et~al.},\ }\href@noop {} {\bibfield  {journal} {\bibinfo  {journal}
  {Nature}\ }\textbf {\bibinfo {volume} {508}},\ \bibinfo {pages} {500}
  (\bibinfo {year} {2014})}\BibitemShut {NoStop}%
\bibitem [{\citenamefont {DiCarlo}\ \emph {et~al.}(2009)\citenamefont
  {DiCarlo}, \citenamefont {Chow}, \citenamefont {Gambetta}, \citenamefont
  {Bishop}, \citenamefont {Johnson}, \citenamefont {Schuster}, \citenamefont
  {Majer}, \citenamefont {Blais}, \citenamefont {Frunzio}, \citenamefont
  {Girvin} \emph {et~al.}}]{Algorithms_Yale1}%
  \BibitemOpen
  \bibfield  {author} {\bibinfo {author} {\bibfnamefont {L.}~\bibnamefont
  {DiCarlo}}, \bibinfo {author} {\bibfnamefont {J.~M.}\ \bibnamefont {Chow}},
  \bibinfo {author} {\bibfnamefont {J.~M.}\ \bibnamefont {Gambetta}}, \bibinfo
  {author} {\bibfnamefont {L.~S.}\ \bibnamefont {Bishop}}, \bibinfo {author}
  {\bibfnamefont {B.~R.}\ \bibnamefont {Johnson}}, \bibinfo {author}
  {\bibfnamefont {D.~I.}\ \bibnamefont {Schuster}}, \bibinfo {author}
  {\bibfnamefont {J.}~\bibnamefont {Majer}}, \bibinfo {author} {\bibfnamefont
  {A.}~\bibnamefont {Blais}}, \bibinfo {author} {\bibfnamefont
  {L.}~\bibnamefont {Frunzio}}, \bibinfo {author} {\bibfnamefont {S.~M.}\
  \bibnamefont {Girvin}},  \emph {et~al.},\ }\href@noop {} {\bibfield
  {journal} {\bibinfo  {journal} {Nature}\ }\textbf {\bibinfo {volume} {460}},\
  \bibinfo {pages} {240} (\bibinfo {year} {2009})}\BibitemShut {NoStop}%
\bibitem [{\citenamefont {Reed}\ \emph {et~al.}(2012)\citenamefont {Reed},
  \citenamefont {DiCarlo}, \citenamefont {Nigg}, \citenamefont {Sun},
  \citenamefont {Frunzio}, \citenamefont {Girvin},\ and\ \citenamefont
  {Schoelkopf}}]{Algorithms_Yale2}%
  \BibitemOpen
  \bibfield  {author} {\bibinfo {author} {\bibfnamefont {M.~D.}\ \bibnamefont
  {Reed}}, \bibinfo {author} {\bibfnamefont {L.}~\bibnamefont {DiCarlo}},
  \bibinfo {author} {\bibfnamefont {S.~E.}\ \bibnamefont {Nigg}}, \bibinfo
  {author} {\bibfnamefont {L.}~\bibnamefont {Sun}}, \bibinfo {author}
  {\bibfnamefont {L.}~\bibnamefont {Frunzio}}, \bibinfo {author} {\bibfnamefont
  {S.~M.}\ \bibnamefont {Girvin}}, \ and\ \bibinfo {author} {\bibfnamefont
  {R.~J.}\ \bibnamefont {Schoelkopf}},\ }\href@noop {} {\bibfield  {journal}
  {\bibinfo  {journal} {Nature}\ }\textbf {\bibinfo {volume} {482}},\ \bibinfo
  {pages} {382} (\bibinfo {year} {2012})}\BibitemShut {NoStop}%
\bibitem [{\citenamefont {Lucero}\ \emph {et~al.}(2012)\citenamefont {Lucero},
  \citenamefont {Barends}, \citenamefont {Chen}, \citenamefont {Kelly},
  \citenamefont {Mariantoni}, \citenamefont {Megrant}, \citenamefont
  {O’Malley}, \citenamefont {Sank}, \citenamefont {Vainsencher},
  \citenamefont {Wenner} \emph {et~al.}}]{Algorithms_UCSB}%
  \BibitemOpen
  \bibfield  {author} {\bibinfo {author} {\bibfnamefont {E.}~\bibnamefont
  {Lucero}}, \bibinfo {author} {\bibfnamefont {R.}~\bibnamefont {Barends}},
  \bibinfo {author} {\bibfnamefont {Y.}~\bibnamefont {Chen}}, \bibinfo {author}
  {\bibfnamefont {J.}~\bibnamefont {Kelly}}, \bibinfo {author} {\bibfnamefont
  {M.}~\bibnamefont {Mariantoni}}, \bibinfo {author} {\bibfnamefont
  {A.}~\bibnamefont {Megrant}}, \bibinfo {author} {\bibfnamefont
  {P.}~\bibnamefont {O’Malley}}, \bibinfo {author} {\bibfnamefont
  {D.}~\bibnamefont {Sank}}, \bibinfo {author} {\bibfnamefont {A.}~\bibnamefont
  {Vainsencher}}, \bibinfo {author} {\bibfnamefont {J.}~\bibnamefont {Wenner}},
   \emph {et~al.},\ }\href@noop {} {\bibfield  {journal} {\bibinfo  {journal}
  {Nat. Phys.}\ }\textbf {\bibinfo {volume} {8}},\ \bibinfo {pages} {719}
  (\bibinfo {year} {2012})}\BibitemShut {NoStop}%
\bibitem [{\citenamefont {Gustavsson}\ \emph {et~al.}(2013)\citenamefont
  {Gustavsson}, \citenamefont {Bylander},\ and\ \citenamefont
  {Oliver}}]{Gustavsson2013a}%
  \BibitemOpen
  \bibfield  {author} {\bibinfo {author} {\bibfnamefont {S.}~\bibnamefont
  {Gustavsson}}, \bibinfo {author} {\bibfnamefont {J.}~\bibnamefont
  {Bylander}}, \ and\ \bibinfo {author} {\bibfnamefont {W.~D.}\ \bibnamefont
  {Oliver}},\ }\href@noop {} {\bibfield  {journal} {\bibinfo  {journal} {Phys.
  Rev. Lett.}\ }\textbf {\bibinfo {volume} {110}},\ \bibinfo {pages} {016603}
  (\bibinfo {year} {2013})}\BibitemShut {NoStop}%
\bibitem [{\citenamefont {Chen}\ \emph {et~al.}(2014)\citenamefont {Chen},
  \citenamefont {Roushan}, \citenamefont {Sank}, \citenamefont {Neill},
  \citenamefont {Lucero} \emph {et~al.}}]{Chen2014a}%
  \BibitemOpen
  \bibfield  {author} {\bibinfo {author} {\bibfnamefont {Y.}~\bibnamefont
  {Chen}}, \bibinfo {author} {\bibfnamefont {P.}~\bibnamefont {Roushan}},
  \bibinfo {author} {\bibfnamefont {D.}~\bibnamefont {Sank}}, \bibinfo {author}
  {\bibfnamefont {C.}~\bibnamefont {Neill}}, \bibinfo {author} {\bibfnamefont
  {E.}~\bibnamefont {Lucero}},  \emph {et~al.},\ }\href {\doibase
  10.1038/ncomms6184} {\bibfield  {journal} {\bibinfo  {journal} {Nat.
  Commun.}\ }\textbf {\bibinfo {volume} {5}},\ \bibinfo {pages} {5184}
  (\bibinfo {year} {2014})}\BibitemShut {NoStop}%
\bibitem [{\citenamefont {Rist{\`e}}\ \emph {et~al.}(2013)\citenamefont
  {Rist{\`e}}, \citenamefont {Dukalski}, \citenamefont {Watson}, \citenamefont
  {de~Lange}, \citenamefont {Tiggelman}, \citenamefont {Blanter}, \citenamefont
  {Lehnert}, \citenamefont {Schouten},\ and\ \citenamefont
  {DiCarlo}}]{Parity_DiCarlo1}%
  \BibitemOpen
  \bibfield  {author} {\bibinfo {author} {\bibfnamefont {D.}~\bibnamefont
  {Rist{\`e}}}, \bibinfo {author} {\bibfnamefont {M.}~\bibnamefont {Dukalski}},
  \bibinfo {author} {\bibfnamefont {C.~A.}\ \bibnamefont {Watson}}, \bibinfo
  {author} {\bibfnamefont {G.}~\bibnamefont {de~Lange}}, \bibinfo {author}
  {\bibfnamefont {M.~J.}\ \bibnamefont {Tiggelman}}, \bibinfo {author}
  {\bibfnamefont {Y.~M.}\ \bibnamefont {Blanter}}, \bibinfo {author}
  {\bibfnamefont {K.~W.}\ \bibnamefont {Lehnert}}, \bibinfo {author}
  {\bibfnamefont {R.~N.}\ \bibnamefont {Schouten}}, \ and\ \bibinfo {author}
  {\bibfnamefont {L.}~\bibnamefont {DiCarlo}},\ }\href@noop {} {\bibfield
  {journal} {\bibinfo  {journal} {Nature}\ }\textbf {\bibinfo {volume} {502}},\
  \bibinfo {pages} {350} (\bibinfo {year} {2013})}\BibitemShut {NoStop}%
\bibitem [{\citenamefont {Saira}\ \emph {et~al.}(2014)\citenamefont {Saira},
  \citenamefont {Groen}, \citenamefont {Cramer}, \citenamefont {Meretska},
  \citenamefont {De~Lange},\ and\ \citenamefont {DiCarlo}}]{Parity_DiCarlo2}%
  \BibitemOpen
  \bibfield  {author} {\bibinfo {author} {\bibfnamefont {O.-P.}\ \bibnamefont
  {Saira}}, \bibinfo {author} {\bibfnamefont {J.~P.}\ \bibnamefont {Groen}},
  \bibinfo {author} {\bibfnamefont {J.}~\bibnamefont {Cramer}}, \bibinfo
  {author} {\bibfnamefont {M.}~\bibnamefont {Meretska}}, \bibinfo {author}
  {\bibfnamefont {G.}~\bibnamefont {De~Lange}}, \ and\ \bibinfo {author}
  {\bibfnamefont {L.}~\bibnamefont {DiCarlo}},\ }\href@noop {} {\bibfield
  {journal} {\bibinfo  {journal} {Phys. Rev. Lett.}\ }\textbf {\bibinfo
  {volume} {112}},\ \bibinfo {pages} {070502} (\bibinfo {year}
  {2014})}\BibitemShut {NoStop}%
\bibitem [{\citenamefont {Chow}\ \emph {et~al.}(2014)\citenamefont {Chow},
  \citenamefont {Gambetta}, \citenamefont {Magesan}, \citenamefont
  {Srinivasan}, \citenamefont {Cross}, \citenamefont {Abraham}, \citenamefont
  {Masluk}, \citenamefont {Johnson}, \citenamefont {Ryan},\ and\ \citenamefont
  {Steffen}}]{Parity_IBM}%
  \BibitemOpen
  \bibfield  {author} {\bibinfo {author} {\bibfnamefont {J.~M.}\ \bibnamefont
  {Chow}}, \bibinfo {author} {\bibfnamefont {J.~M.}\ \bibnamefont {Gambetta}},
  \bibinfo {author} {\bibfnamefont {E.}~\bibnamefont {Magesan}}, \bibinfo
  {author} {\bibfnamefont {S.~J.}\ \bibnamefont {Srinivasan}}, \bibinfo
  {author} {\bibfnamefont {A.~W.}\ \bibnamefont {Cross}}, \bibinfo {author}
  {\bibfnamefont {D.~W.}\ \bibnamefont {Abraham}}, \bibinfo {author}
  {\bibfnamefont {N.~A.}\ \bibnamefont {Masluk}}, \bibinfo {author}
  {\bibfnamefont {B.}~\bibnamefont {Johnson}}, \bibinfo {author} {\bibfnamefont
  {C.~A.}\ \bibnamefont {Ryan}}, \ and\ \bibinfo {author} {\bibfnamefont
  {M.}~\bibnamefont {Steffen}},\ }\href@noop {} {\bibfield  {journal} {\bibinfo
   {journal} {Nat. Commun.}\ }\textbf {\bibinfo {volume} {5}},\ \bibinfo
  {pages} {4015} (\bibinfo {year} {2014})}\BibitemShut {NoStop}%
\bibitem [{\citenamefont {Kelly}\ \emph {et~al.}(2015)\citenamefont {Kelly},
  \citenamefont {Barends}, \citenamefont {Fowler}, \citenamefont {Megrant},
  \citenamefont {Jeffrey}, \citenamefont {White}, \citenamefont {Sank},
  \citenamefont {Mutus}, \citenamefont {Campbell}, \citenamefont {Chen},
  \citenamefont {Chen}, \citenamefont {Chiaro}, \citenamefont {Dunsworth},
  \citenamefont {Hoi}, \citenamefont {Neill}, \citenamefont {O'Malley},
  \citenamefont {Quintana}, \citenamefont {Roushan}, \citenamefont
  {Vainsencher}, \citenamefont {Wenner}, \citenamefont {Cleland},\ and\
  \citenamefont {Martinis}}]{Kelly15a_UCSB_Nature}%
  \BibitemOpen
  \bibfield  {author} {\bibinfo {author} {\bibfnamefont {J.}~\bibnamefont
  {Kelly}}, \bibinfo {author} {\bibfnamefont {R.}~\bibnamefont {Barends}},
  \bibinfo {author} {\bibfnamefont {A.~G.}\ \bibnamefont {Fowler}}, \bibinfo
  {author} {\bibfnamefont {A.}~\bibnamefont {Megrant}}, \bibinfo {author}
  {\bibfnamefont {E.}~\bibnamefont {Jeffrey}}, \bibinfo {author} {\bibfnamefont
  {T.~C.}\ \bibnamefont {White}}, \bibinfo {author} {\bibfnamefont
  {D.}~\bibnamefont {Sank}}, \bibinfo {author} {\bibfnamefont {J.~Y.}\
  \bibnamefont {Mutus}}, \bibinfo {author} {\bibfnamefont {B.}~\bibnamefont
  {Campbell}}, \bibinfo {author} {\bibfnamefont {Y.}~\bibnamefont {Chen}},
  \bibinfo {author} {\bibfnamefont {Z.}~\bibnamefont {Chen}}, \bibinfo {author}
  {\bibfnamefont {B.}~\bibnamefont {Chiaro}}, \bibinfo {author} {\bibfnamefont
  {A.}~\bibnamefont {Dunsworth}}, \bibinfo {author} {\bibfnamefont {I.~C.}\
  \bibnamefont {Hoi}}, \bibinfo {author} {\bibfnamefont {C.}~\bibnamefont
  {Neill}}, \bibinfo {author} {\bibfnamefont {P.~J.~J.}\ \bibnamefont
  {O'Malley}}, \bibinfo {author} {\bibfnamefont {C.}~\bibnamefont {Quintana}},
  \bibinfo {author} {\bibfnamefont {P.}~\bibnamefont {Roushan}}, \bibinfo
  {author} {\bibfnamefont {A.}~\bibnamefont {Vainsencher}}, \bibinfo {author}
  {\bibfnamefont {J.}~\bibnamefont {Wenner}}, \bibinfo {author} {\bibfnamefont
  {A.~N.}\ \bibnamefont {Cleland}}, \ and\ \bibinfo {author} {\bibfnamefont
  {J.~M.}\ \bibnamefont {Martinis}},\ }\href {\doibase 10.1038/nature14270}
  {\bibfield  {journal} {\bibinfo  {journal} {Nature}\ }\textbf {\bibinfo
  {volume} {519}},\ \bibinfo {pages} {66} (\bibinfo {year} {2015})}\BibitemShut
  {NoStop}%
\bibitem [{\citenamefont {C{\'o}rcoles}\ \emph {et~al.}(2014)\citenamefont
  {C{\'o}rcoles}, \citenamefont {Magesan}, \citenamefont {Srinivasan},
  \citenamefont {Cross}, \citenamefont {Steffen}, \citenamefont {Gambetta},\
  and\ \citenamefont {Chow}}]{Corcoles15a_arXiv}%
  \BibitemOpen
  \bibfield  {author} {\bibinfo {author} {\bibfnamefont {A.}~\bibnamefont
  {C{\'o}rcoles}}, \bibinfo {author} {\bibfnamefont {E.}~\bibnamefont
  {Magesan}}, \bibinfo {author} {\bibfnamefont {S.~J.}\ \bibnamefont
  {Srinivasan}}, \bibinfo {author} {\bibfnamefont {A.~W.}\ \bibnamefont
  {Cross}}, \bibinfo {author} {\bibfnamefont {M.}~\bibnamefont {Steffen}},
  \bibinfo {author} {\bibfnamefont {J.~M.}\ \bibnamefont {Gambetta}}, \ and\
  \bibinfo {author} {\bibfnamefont {J.~M.}\ \bibnamefont {Chow}},\ }\href@noop
  {} {\bibfield  {journal} {\bibinfo  {journal} {arXiv:1410.6419}\ } (\bibinfo
  {year} {2014})}\BibitemShut {NoStop}%
\bibitem [{\citenamefont {Rist{\`e}}\ \emph {et~al.}(2015)\citenamefont
  {Rist{\`e}}, \citenamefont {Poletto}, \citenamefont {Huang}, \citenamefont
  {Bruno}, \citenamefont {Vesterinen}, \citenamefont {Saira},\ and\
  \citenamefont {DiCarlo}}]{Riste15a_logical_qubit_NatComm}%
  \BibitemOpen
  \bibfield  {author} {\bibinfo {author} {\bibfnamefont {D.}~\bibnamefont
  {Rist{\`e}}}, \bibinfo {author} {\bibfnamefont {S.}~\bibnamefont {Poletto}},
  \bibinfo {author} {\bibfnamefont {M.-Z.}\ \bibnamefont {Huang}}, \bibinfo
  {author} {\bibfnamefont {A.}~\bibnamefont {Bruno}}, \bibinfo {author}
  {\bibfnamefont {V.}~\bibnamefont {Vesterinen}}, \bibinfo {author}
  {\bibfnamefont {O.-P.}\ \bibnamefont {Saira}}, \ and\ \bibinfo {author}
  {\bibfnamefont {L.}~\bibnamefont {DiCarlo}},\ }\href {\doibase
  10.1038/ncomms7983} {\bibfield  {journal} {\bibinfo  {journal} {Nat.
  Commun.}\ }\textbf {\bibinfo {volume} {6}},\ \bibinfo {pages} {6983}
  (\bibinfo {year} {2015})}\BibitemShut {NoStop}%
\bibitem [{\citenamefont {Sun}\ \emph {et~al.}(2014)\citenamefont {Sun},
  \citenamefont {Petrenko}, \citenamefont {Leghtas}, \citenamefont {Vlastakis},
  \citenamefont {Kirchmair}, \citenamefont {Sliwa}, \citenamefont {Narla},
  \citenamefont {Hatridge}, \citenamefont {Shankar}, \citenamefont {Blumoff}
  \emph {et~al.}}]{Parity_Yale}%
  \BibitemOpen
  \bibfield  {author} {\bibinfo {author} {\bibfnamefont {L.}~\bibnamefont
  {Sun}}, \bibinfo {author} {\bibfnamefont {A.}~\bibnamefont {Petrenko}},
  \bibinfo {author} {\bibfnamefont {Z.}~\bibnamefont {Leghtas}}, \bibinfo
  {author} {\bibfnamefont {B.}~\bibnamefont {Vlastakis}}, \bibinfo {author}
  {\bibfnamefont {G.}~\bibnamefont {Kirchmair}}, \bibinfo {author}
  {\bibfnamefont {K.~M.}\ \bibnamefont {Sliwa}}, \bibinfo {author}
  {\bibfnamefont {A.}~\bibnamefont {Narla}}, \bibinfo {author} {\bibfnamefont
  {M.}~\bibnamefont {Hatridge}}, \bibinfo {author} {\bibfnamefont
  {S.}~\bibnamefont {Shankar}}, \bibinfo {author} {\bibfnamefont
  {J.}~\bibnamefont {Blumoff}},  \emph {et~al.},\ }\href@noop {} {\bibfield
  {journal} {\bibinfo  {journal} {Nature}\ }\textbf {\bibinfo {volume} {511}},\
  \bibinfo {pages} {444} (\bibinfo {year} {2014})}\BibitemShut {NoStop}%
\bibitem [{\citenamefont {Paik}\ \emph {et~al.}(2011)\citenamefont {Paik},
  \citenamefont {Schuster}, \citenamefont {Bishop}, \citenamefont {Kirchmair},
  \citenamefont {Catelani}, \citenamefont {Sears}, \citenamefont {Johnson},
  \citenamefont {Reagor}, \citenamefont {Funzio}, \citenamefont {Glazman},
  \citenamefont {Girvin}, \citenamefont {Devoret},\ and\ \citenamefont
  {Schoelkopf}}]{Paik2011a}%
  \BibitemOpen
  \bibfield  {author} {\bibinfo {author} {\bibfnamefont {H.}~\bibnamefont
  {Paik}}, \bibinfo {author} {\bibfnamefont {D.~I.}\ \bibnamefont {Schuster}},
  \bibinfo {author} {\bibfnamefont {L.~S.}\ \bibnamefont {Bishop}}, \bibinfo
  {author} {\bibfnamefont {G.}~\bibnamefont {Kirchmair}}, \bibinfo {author}
  {\bibfnamefont {G.}~\bibnamefont {Catelani}}, \bibinfo {author}
  {\bibfnamefont {A.~P.}\ \bibnamefont {Sears}}, \bibinfo {author}
  {\bibfnamefont {B.~R.}\ \bibnamefont {Johnson}}, \bibinfo {author}
  {\bibfnamefont {M.~J.}\ \bibnamefont {Reagor}}, \bibinfo {author}
  {\bibfnamefont {L.}~\bibnamefont {Funzio}}, \bibinfo {author} {\bibfnamefont
  {L.~I.}\ \bibnamefont {Glazman}}, \bibinfo {author} {\bibfnamefont {S.~M.}\
  \bibnamefont {Girvin}}, \bibinfo {author} {\bibfnamefont {M.~H.}\
  \bibnamefont {Devoret}}, \ and\ \bibinfo {author} {\bibfnamefont {R.~J.}\
  \bibnamefont {Schoelkopf}},\ }\href@noop {} {\bibfield  {journal} {\bibinfo
  {journal} {Phys. Rev. Lett.}\ }\textbf {\bibinfo {volume} {107}},\ \bibinfo
  {pages} {240501} (\bibinfo {year} {2011})}\BibitemShut {NoStop}%
\bibitem [{\citenamefont {Corcoles}\ \emph {et~al.}(2011)\citenamefont
  {Corcoles}, \citenamefont {Chow}, \citenamefont {Gambetta}, \citenamefont
  {Rigetti}, \citenamefont {Rozen}, \citenamefont {Keefe}, \citenamefont
  {Rothwell}, \citenamefont {Ketchen},\ and\ \citenamefont
  {Steffen}}]{IBM_Temperature}%
  \BibitemOpen
  \bibfield  {author} {\bibinfo {author} {\bibfnamefont {A.~D.}\ \bibnamefont
  {Corcoles}}, \bibinfo {author} {\bibfnamefont {J.~M.}\ \bibnamefont {Chow}},
  \bibinfo {author} {\bibfnamefont {J.~M.}\ \bibnamefont {Gambetta}}, \bibinfo
  {author} {\bibfnamefont {C.}~\bibnamefont {Rigetti}}, \bibinfo {author}
  {\bibfnamefont {J.}~\bibnamefont {Rozen}}, \bibinfo {author} {\bibfnamefont
  {G.~A.}\ \bibnamefont {Keefe}}, \bibinfo {author} {\bibfnamefont {M.~B.}\
  \bibnamefont {Rothwell}}, \bibinfo {author} {\bibfnamefont {M.~B.}\
  \bibnamefont {Ketchen}}, \ and\ \bibinfo {author} {\bibfnamefont
  {M.}~\bibnamefont {Steffen}},\ }\href@noop {} {\bibfield  {journal} {\bibinfo
   {journal} {Appl. Phys. Lett.}\ }\textbf {\bibinfo {volume} {99}},\ \bibinfo
  {pages} {181906} (\bibinfo {year} {2011})}\BibitemShut {NoStop}%
\bibitem [{\citenamefont {Johnson}\ \emph {et~al.}(2012)\citenamefont
  {Johnson}, \citenamefont {Macklin}, \citenamefont {Slichter}, \citenamefont
  {Vijay}, \citenamefont {Weingarten}, \citenamefont {Clarke},\ and\
  \citenamefont {Siddiqi}}]{Berkeley_Temperature}%
  \BibitemOpen
  \bibfield  {author} {\bibinfo {author} {\bibfnamefont {J.~E.}\ \bibnamefont
  {Johnson}}, \bibinfo {author} {\bibfnamefont {C.}~\bibnamefont {Macklin}},
  \bibinfo {author} {\bibfnamefont {D.~H.}\ \bibnamefont {Slichter}}, \bibinfo
  {author} {\bibfnamefont {R.}~\bibnamefont {Vijay}}, \bibinfo {author}
  {\bibfnamefont {E.~B.}\ \bibnamefont {Weingarten}}, \bibinfo {author}
  {\bibfnamefont {J.}~\bibnamefont {Clarke}}, \ and\ \bibinfo {author}
  {\bibfnamefont {I.}~\bibnamefont {Siddiqi}},\ }\href@noop {} {\bibfield
  {journal} {\bibinfo  {journal} {Phys. Rev. Lett.}\ }\textbf {\bibinfo
  {volume} {109}},\ \bibinfo {pages} {050506} (\bibinfo {year}
  {2012})}\BibitemShut {NoStop}%
\bibitem [{\citenamefont {Rist{\`e}}\ \emph {et~al.}(2012)\citenamefont
  {Rist{\`e}}, \citenamefont {Bultink}, \citenamefont {Lehnert},\ and\
  \citenamefont {DiCarlo}}]{Delft_Temperature}%
  \BibitemOpen
  \bibfield  {author} {\bibinfo {author} {\bibfnamefont {D.}~\bibnamefont
  {Rist{\`e}}}, \bibinfo {author} {\bibfnamefont {C.~C.}\ \bibnamefont
  {Bultink}}, \bibinfo {author} {\bibfnamefont {K.~W.}\ \bibnamefont
  {Lehnert}}, \ and\ \bibinfo {author} {\bibfnamefont {L.}~\bibnamefont
  {DiCarlo}},\ }\href@noop {} {\bibfield  {journal} {\bibinfo  {journal} {Phys.
  Rev. Lett.}\ }\textbf {\bibinfo {volume} {109}},\ \bibinfo {pages} {240502}
  (\bibinfo {year} {2012})}\BibitemShut {NoStop}%
\bibitem [{\citenamefont {Giazotto}\ \emph {et~al.}(2006)\citenamefont
  {Giazotto}, \citenamefont {Heikkil{\"a}}, \citenamefont {Luukanen},
  \citenamefont {Savin},\ and\ \citenamefont {Pekola}}]{Thermal1}%
  \BibitemOpen
  \bibfield  {author} {\bibinfo {author} {\bibfnamefont {F.}~\bibnamefont
  {Giazotto}}, \bibinfo {author} {\bibfnamefont {T.~T.}\ \bibnamefont
  {Heikkil{\"a}}}, \bibinfo {author} {\bibfnamefont {A.}~\bibnamefont
  {Luukanen}}, \bibinfo {author} {\bibfnamefont {A.~M.}\ \bibnamefont {Savin}},
  \ and\ \bibinfo {author} {\bibfnamefont {J.~P.}\ \bibnamefont {Pekola}},\
  }\href@noop {} {\bibfield  {journal} {\bibinfo  {journal} {Reviews of Modern
  Physics}\ }\textbf {\bibinfo {volume} {78}},\ \bibinfo {pages} {217}
  (\bibinfo {year} {2006})}\BibitemShut {NoStop}%
\bibitem [{\citenamefont {Bladh}\ \emph {et~al.}(2003)\citenamefont {Bladh},
  \citenamefont {Gunnarsson}, \citenamefont {H\"{u}rfeld}, \citenamefont
  {Devi}, \citenamefont {Kristoffersson}, \citenamefont {Sm{\aa}lander},
  \citenamefont {Pehrson}, \citenamefont {Cleason}, \citenamefont {Delsing},\
  and\ \citenamefont {Taslakov}}]{Bladh03a}%
  \BibitemOpen
  \bibfield  {author} {\bibinfo {author} {\bibfnamefont {K.}~\bibnamefont
  {Bladh}}, \bibinfo {author} {\bibfnamefont {D.}~\bibnamefont {Gunnarsson}},
  \bibinfo {author} {\bibfnamefont {E.}~\bibnamefont {H\"{u}rfeld}}, \bibinfo
  {author} {\bibfnamefont {S.}~\bibnamefont {Devi}}, \bibinfo {author}
  {\bibfnamefont {C.}~\bibnamefont {Kristoffersson}}, \bibinfo {author}
  {\bibfnamefont {B.}~\bibnamefont {Sm{\aa}lander}}, \bibinfo {author}
  {\bibfnamefont {S.}~\bibnamefont {Pehrson}}, \bibinfo {author} {\bibfnamefont
  {T.}~\bibnamefont {Cleason}}, \bibinfo {author} {\bibfnamefont
  {P.}~\bibnamefont {Delsing}}, \ and\ \bibinfo {author} {\bibfnamefont
  {M.}~\bibnamefont {Taslakov}},\ }\href@noop {} {\bibfield  {journal}
  {\bibinfo  {journal} {Rev. Sci. Inst.}\ }\textbf {\bibinfo {volume} {74}},\
  \bibinfo {pages} {1323} (\bibinfo {year} {2003})}\BibitemShut {NoStop}%
\bibitem [{\citenamefont {Vion}\ \emph {et~al.}(1995)\citenamefont {Vion},
  \citenamefont {Orfila}, \citenamefont {Joyez}, \citenamefont {Esteve},\ and\
  \citenamefont {Devoret}}]{Vion1995a}%
  \BibitemOpen
  \bibfield  {author} {\bibinfo {author} {\bibfnamefont {D.}~\bibnamefont
  {Vion}}, \bibinfo {author} {\bibfnamefont {P.~F.}\ \bibnamefont {Orfila}},
  \bibinfo {author} {\bibfnamefont {P.}~\bibnamefont {Joyez}}, \bibinfo
  {author} {\bibfnamefont {D.}~\bibnamefont {Esteve}}, \ and\ \bibinfo {author}
  {\bibfnamefont {M.~H.}\ \bibnamefont {Devoret}},\ }\href@noop {} {\bibfield
  {journal} {\bibinfo  {journal} {Phys. Rev. Lett.}\ }\textbf {\bibinfo
  {volume} {77}},\ \bibinfo {pages} {2519} (\bibinfo {year}
  {1995})}\BibitemShut {NoStop}%
\bibitem [{\citenamefont {le~Sueur}\ and\ \citenamefont
  {Joyez}(2006)}]{Sueur06a}%
  \BibitemOpen
  \bibfield  {author} {\bibinfo {author} {\bibfnamefont {H.}~\bibnamefont
  {le~Sueur}}\ and\ \bibinfo {author} {\bibfnamefont {P.}~\bibnamefont
  {Joyez}},\ }\href@noop {} {\bibfield  {journal} {\bibinfo  {journal} {Rev.
  Sci. Inst.}\ }\textbf {\bibinfo {volume} {77}},\ \bibinfo {pages} {115102}
  (\bibinfo {year} {2006})}\BibitemShut {NoStop}%
\bibitem [{\citenamefont {Devoret}\ \emph {et~al.}(1985)\citenamefont
  {Devoret}, \citenamefont {Martinis},\ and\ \citenamefont
  {Clarke}}]{Devoret1985a}%
  \BibitemOpen
  \bibfield  {author} {\bibinfo {author} {\bibfnamefont {M.~H.}\ \bibnamefont
  {Devoret}}, \bibinfo {author} {\bibfnamefont {J.}~\bibnamefont {Martinis}}, \
  and\ \bibinfo {author} {\bibfnamefont {J.}~\bibnamefont {Clarke}},\
  }\href@noop {} {\bibfield  {journal} {\bibinfo  {journal} {Phys. Rev. Lett.}\
  }\textbf {\bibinfo {volume} {55}},\ \bibinfo {pages} {1543} (\bibinfo {year}
  {1985})}\BibitemShut {NoStop}%
\bibitem [{\citenamefont {Martinis}\ \emph {et~al.}(1987)\citenamefont
  {Martinis}, \citenamefont {Devoret},\ and\ \citenamefont
  {Clarke}}]{Martinis1987a}%
  \BibitemOpen
  \bibfield  {author} {\bibinfo {author} {\bibfnamefont {J.}~\bibnamefont
  {Martinis}}, \bibinfo {author} {\bibfnamefont {M.~H.}\ \bibnamefont
  {Devoret}}, \ and\ \bibinfo {author} {\bibfnamefont {J.}~\bibnamefont
  {Clarke}},\ }\href@noop {} {\bibfield  {journal} {\bibinfo  {journal} {Phys.
  Rev. B}\ }\textbf {\bibinfo {volume} {35}},\ \bibinfo {pages} {4682}
  (\bibinfo {year} {1987})}\BibitemShut {NoStop}%
\bibitem [{\citenamefont {Milliken}\ \emph {et~al.}(2007)\citenamefont
  {Milliken}, \citenamefont {Rozen}, \citenamefont {Keefe},\ and\ \citenamefont
  {Koch}}]{Milliken07a}%
  \BibitemOpen
  \bibfield  {author} {\bibinfo {author} {\bibfnamefont {F.~P.}\ \bibnamefont
  {Milliken}}, \bibinfo {author} {\bibfnamefont {J.~R.}\ \bibnamefont {Rozen}},
  \bibinfo {author} {\bibfnamefont {G.~A.}\ \bibnamefont {Keefe}}, \ and\
  \bibinfo {author} {\bibfnamefont {R.~H.}\ \bibnamefont {Koch}},\ }\href@noop
  {} {\bibfield  {journal} {\bibinfo  {journal} {Rev. Sci. Inst.}\ }\textbf
  {\bibinfo {volume} {78}},\ \bibinfo {pages} {024701} (\bibinfo {year}
  {2007})}\BibitemShut {NoStop}%
\bibitem [{\citenamefont {Lukashenko}\ and\ \citenamefont
  {Ustinov}(2008)}]{Lukashenko08a}%
  \BibitemOpen
  \bibfield  {author} {\bibinfo {author} {\bibfnamefont {A.}~\bibnamefont
  {Lukashenko}}\ and\ \bibinfo {author} {\bibfnamefont {A.}~\bibnamefont
  {Ustinov}},\ }\href@noop {} {\bibfield  {journal} {\bibinfo  {journal} {Rev.
  Sci. Inst.}\ }\textbf {\bibinfo {volume} {79}},\ \bibinfo {pages} {014701}
  (\bibinfo {year} {2008})}\BibitemShut {NoStop}%
\bibitem [{\citenamefont {Fukushima}\ \emph {et~al.}(1997)\citenamefont
  {Fukushima}, \citenamefont {Sato}, \citenamefont {Iwasa}, \citenamefont
  {Nakamura}, \citenamefont {Komatsuzaki},\ and\ \citenamefont
  {Sakamoto}}]{Fukushima97a}%
  \BibitemOpen
  \bibfield  {author} {\bibinfo {author} {\bibfnamefont {A.}~\bibnamefont
  {Fukushima}}, \bibinfo {author} {\bibfnamefont {A.}~\bibnamefont {Sato}},
  \bibinfo {author} {\bibfnamefont {A.}~\bibnamefont {Iwasa}}, \bibinfo
  {author} {\bibfnamefont {Y.}~\bibnamefont {Nakamura}}, \bibinfo {author}
  {\bibfnamefont {Y.}~\bibnamefont {Komatsuzaki}}, \ and\ \bibinfo {author}
  {\bibfnamefont {Y.}~\bibnamefont {Sakamoto}},\ }\href@noop {} {\bibfield
  {journal} {\bibinfo  {journal} {IEEE Trans. Instrum. Meas.}\ }\textbf
  {\bibinfo {volume} {46}},\ \bibinfo {pages} {289} (\bibinfo {year}
  {1997})}\BibitemShut {NoStop}%
\bibitem [{\citenamefont {Zorin}(1995)}]{Zorin1995a}%
  \BibitemOpen
  \bibfield  {author} {\bibinfo {author} {\bibfnamefont {A.~B.}\ \bibnamefont
  {Zorin}},\ }\href@noop {} {\bibfield  {journal} {\bibinfo  {journal} {Rev.
  Sci. Inst.}\ }\textbf {\bibinfo {volume} {66}},\ \bibinfo {pages} {4296}
  (\bibinfo {year} {1995})}\BibitemShut {NoStop}%
\bibitem [{\citenamefont {Glattli}\ \emph {et~al.}(1997)\citenamefont
  {Glattli}, \citenamefont {Jacques}, \citenamefont {Kumar}, \citenamefont
  {Pari},\ and\ \citenamefont {Saminadayar}}]{Glattli1997a}%
  \BibitemOpen
  \bibfield  {author} {\bibinfo {author} {\bibfnamefont {D.~C.}\ \bibnamefont
  {Glattli}}, \bibinfo {author} {\bibfnamefont {P.}~\bibnamefont {Jacques}},
  \bibinfo {author} {\bibfnamefont {A.}~\bibnamefont {Kumar}}, \bibinfo
  {author} {\bibfnamefont {P.}~\bibnamefont {Pari}}, \ and\ \bibinfo {author}
  {\bibfnamefont {L.}~\bibnamefont {Saminadayar}},\ }\href@noop {} {\bibfield
  {journal} {\bibinfo  {journal} {J. Appl. Phys.}\ }\textbf {\bibinfo {volume}
  {81}},\ \bibinfo {pages} {7350} (\bibinfo {year} {1997})}\BibitemShut
  {NoStop}%
\bibitem [{\citenamefont {Courtois}\ \emph {et~al.}(1995)\citenamefont
  {Courtois}, \citenamefont {Buisson}, \citenamefont {Chaussy},\ and\
  \citenamefont {Pannetier}}]{Courtois95a}%
  \BibitemOpen
  \bibfield  {author} {\bibinfo {author} {\bibfnamefont {H.}~\bibnamefont
  {Courtois}}, \bibinfo {author} {\bibfnamefont {O.}~\bibnamefont {Buisson}},
  \bibinfo {author} {\bibfnamefont {J.}~\bibnamefont {Chaussy}}, \ and\
  \bibinfo {author} {\bibfnamefont {B.}~\bibnamefont {Pannetier}},\ }\href@noop
  {} {\bibfield  {journal} {\bibinfo  {journal} {Rev. Sci. Inst.}\ }\textbf
  {\bibinfo {volume} {66}},\ \bibinfo {pages} {3465} (\bibinfo {year}
  {1995})}\BibitemShut {NoStop}%
\bibitem [{\citenamefont {Santavicca}\ and\ \citenamefont
  {Prober}(2008)}]{Santavicca08a}%
  \BibitemOpen
  \bibfield  {author} {\bibinfo {author} {\bibfnamefont {D.~F.}\ \bibnamefont
  {Santavicca}}\ and\ \bibinfo {author} {\bibfnamefont {D.~E.}\ \bibnamefont
  {Prober}},\ }\href@noop {} {\bibfield  {journal} {\bibinfo  {journal} {Meas.
  Sci. Technol.}\ }\textbf {\bibinfo {volume} {19}},\ \bibinfo {pages} {087001}
  (\bibinfo {year} {2008})}\BibitemShut {NoStop}%
\bibitem [{\citenamefont {Slichter}\ \emph {et~al.}(2009)\citenamefont
  {Slichter}, \citenamefont {Naaman},\ and\ \citenamefont
  {Siddiqi}}]{Slichter09a}%
  \BibitemOpen
  \bibfield  {author} {\bibinfo {author} {\bibfnamefont {D.~H.}\ \bibnamefont
  {Slichter}}, \bibinfo {author} {\bibfnamefont {O.}~\bibnamefont {Naaman}}, \
  and\ \bibinfo {author} {\bibfnamefont {I.}~\bibnamefont {Siddiqi}},\
  }\href@noop {} {\bibfield  {journal} {\bibinfo  {journal} {Appl. Phys.
  Lett.}\ }\textbf {\bibinfo {volume} {94}},\ \bibinfo {pages} {192508}
  (\bibinfo {year} {2009})}\BibitemShut {NoStop}%
\bibitem [{\citenamefont {Vijay}\ \emph {et~al.}(2009)\citenamefont {Vijay},
  \citenamefont {Devoret},\ and\ \citenamefont {Siddiqi}}]{Vijay09a}%
  \BibitemOpen
  \bibfield  {author} {\bibinfo {author} {\bibfnamefont {R.}~\bibnamefont
  {Vijay}}, \bibinfo {author} {\bibfnamefont {M.~H.}\ \bibnamefont {Devoret}},
  \ and\ \bibinfo {author} {\bibfnamefont {I.}~\bibnamefont {Siddiqi}},\
  }\href@noop {} {\bibfield  {journal} {\bibinfo  {journal} {Rev. Sci. Inst.}\
  }\textbf {\bibinfo {volume} {80}},\ \bibinfo {pages} {111101} (\bibinfo
  {year} {2009})}\BibitemShut {NoStop}%
\bibitem [{\citenamefont {Hergenrother}\ \emph {et~al.}(1995)\citenamefont
  {Hergenrother}, \citenamefont {Lu}, \citenamefont {Tuominen}, \citenamefont
  {Ralph},\ and\ \citenamefont {Tinkham}}]{Hergenrother1995a}%
  \BibitemOpen
  \bibfield  {author} {\bibinfo {author} {\bibfnamefont {J.~M.}\ \bibnamefont
  {Hergenrother}}, \bibinfo {author} {\bibfnamefont {J.~G.}\ \bibnamefont
  {Lu}}, \bibinfo {author} {\bibfnamefont {M.~T.}\ \bibnamefont {Tuominen}},
  \bibinfo {author} {\bibfnamefont {D.~C.}\ \bibnamefont {Ralph}}, \ and\
  \bibinfo {author} {\bibfnamefont {M.}~\bibnamefont {Tinkham}},\ }\href@noop
  {} {\bibfield  {journal} {\bibinfo  {journal} {Phys. Rev. B}\ }\textbf
  {\bibinfo {volume} {51}},\ \bibinfo {pages} {9407} (\bibinfo {year}
  {1995})}\BibitemShut {NoStop}%
\bibitem [{\citenamefont {Persky}(1999)}]{Persky99a}%
  \BibitemOpen
  \bibfield  {author} {\bibinfo {author} {\bibfnamefont {M.}~\bibnamefont
  {Persky}},\ }\href@noop {} {\bibfield  {journal} {\bibinfo  {journal} {Rev.
  Sci. Inst.}\ }\textbf {\bibinfo {volume} {70}},\ \bibinfo {pages} {2193}
  (\bibinfo {year} {1999})}\BibitemShut {NoStop}%
\bibitem [{\citenamefont {Barends}\ \emph {et~al.}(2011)\citenamefont
  {Barends}, \citenamefont {Wenner}, \citenamefont {Lenander}, \citenamefont
  {Chen}, \citenamefont {Bialczak}, \citenamefont {Kelly}, \citenamefont
  {Lucero}, \citenamefont {O’Malley}, \citenamefont {Mariantoni},
  \citenamefont {Sank} \emph {et~al.}}]{UCSB_Wiring}%
  \BibitemOpen
  \bibfield  {author} {\bibinfo {author} {\bibfnamefont {R.}~\bibnamefont
  {Barends}}, \bibinfo {author} {\bibfnamefont {J.}~\bibnamefont {Wenner}},
  \bibinfo {author} {\bibfnamefont {M.}~\bibnamefont {Lenander}}, \bibinfo
  {author} {\bibfnamefont {Y.}~\bibnamefont {Chen}}, \bibinfo {author}
  {\bibfnamefont {R.~C.}\ \bibnamefont {Bialczak}}, \bibinfo {author}
  {\bibfnamefont {J.}~\bibnamefont {Kelly}}, \bibinfo {author} {\bibfnamefont
  {E.}~\bibnamefont {Lucero}}, \bibinfo {author} {\bibfnamefont
  {P.}~\bibnamefont {O’Malley}}, \bibinfo {author} {\bibfnamefont
  {M.}~\bibnamefont {Mariantoni}}, \bibinfo {author} {\bibfnamefont
  {D.}~\bibnamefont {Sank}},  \emph {et~al.},\ }\href@noop {} {\bibfield
  {journal} {\bibinfo  {journal} {Appl. Phys. Lett.}\ }\textbf {\bibinfo
  {volume} {99}},\ \bibinfo {pages} {113507} (\bibinfo {year}
  {2011})}\BibitemShut {NoStop}%
\bibitem [{Sup()}]{Supplementary}%
  \BibitemOpen
  \href@noop {} {\bibinfo  {journal} {See Supplementary Material at [URL],
  which includes Refs. \cite{Yan2014a, Ambegaokar63a, Ambegaokar63b}}\
  }\BibitemShut {NoStop}%
\bibitem [{\citenamefont {Peterer}\ \emph {et~al.}(2015)\citenamefont
  {Peterer}, \citenamefont {Bader}, \citenamefont {Jin}, \citenamefont {Yan},
  \citenamefont {Kamal}, \citenamefont {Gudmundsen}, \citenamefont {Leek},
  \citenamefont {Orlando}, \citenamefont {Oliver},\ and\ \citenamefont
  {Gustavsson}}]{Peterer15a_PRL}%
  \BibitemOpen
\bibfield  {journal} {  }\bibfield  {author} {\bibinfo {author} {\bibfnamefont
  {M.~J.}\ \bibnamefont {Peterer}}, \bibinfo {author} {\bibfnamefont {S.~J.}\
  \bibnamefont {Bader}}, \bibinfo {author} {\bibfnamefont {X.}~\bibnamefont
  {Jin}}, \bibinfo {author} {\bibfnamefont {F.}~\bibnamefont {Yan}}, \bibinfo
  {author} {\bibfnamefont {A.}~\bibnamefont {Kamal}}, \bibinfo {author}
  {\bibfnamefont {T.~J.}\ \bibnamefont {Gudmundsen}}, \bibinfo {author}
  {\bibfnamefont {P.~J.}\ \bibnamefont {Leek}}, \bibinfo {author}
  {\bibfnamefont {T.~P.}\ \bibnamefont {Orlando}}, \bibinfo {author}
  {\bibfnamefont {W.~D.}\ \bibnamefont {Oliver}}, \ and\ \bibinfo {author}
  {\bibfnamefont {S.}~\bibnamefont {Gustavsson}},\ }\href {\doibase
  10.1103/PhysRevLett.114.010501} {\bibfield  {journal} {\bibinfo  {journal}
  {Phys. Rev. Lett.}\ }\textbf {\bibinfo {volume} {114}},\ \bibinfo {pages}
  {010501} (\bibinfo {year} {2015})}\BibitemShut {NoStop}%
\bibitem [{\citenamefont {Wenner}\ \emph {et~al.}(2013)\citenamefont {Wenner},
  \citenamefont {Yin}, \citenamefont {Lucero}, \citenamefont {Barends},
  \citenamefont {Chen}, \citenamefont {Chiaro}, \citenamefont {Kelly},
  \citenamefont {Lenander}, \citenamefont {Mariantoni}, \citenamefont
  {Megrant}, \citenamefont {Neill}, \citenamefont {O'Malley}, \citenamefont
  {Sank}, \citenamefont {Vainsencher}, \citenamefont {Wang}, \citenamefont
  {White}, \citenamefont {Cleland},\ and\ \citenamefont
  {Martinis}}]{UCSB_Hot_QP}%
  \BibitemOpen
  \bibfield  {author} {\bibinfo {author} {\bibfnamefont {J.}~\bibnamefont
  {Wenner}}, \bibinfo {author} {\bibfnamefont {Y.}~\bibnamefont {Yin}},
  \bibinfo {author} {\bibfnamefont {E.}~\bibnamefont {Lucero}}, \bibinfo
  {author} {\bibfnamefont {R.}~\bibnamefont {Barends}}, \bibinfo {author}
  {\bibfnamefont {Y.}~\bibnamefont {Chen}}, \bibinfo {author} {\bibfnamefont
  {B.}~\bibnamefont {Chiaro}}, \bibinfo {author} {\bibfnamefont
  {J.}~\bibnamefont {Kelly}}, \bibinfo {author} {\bibfnamefont
  {M.}~\bibnamefont {Lenander}}, \bibinfo {author} {\bibfnamefont
  {M.}~\bibnamefont {Mariantoni}}, \bibinfo {author} {\bibfnamefont
  {A.}~\bibnamefont {Megrant}}, \bibinfo {author} {\bibfnamefont
  {C.}~\bibnamefont {Neill}}, \bibinfo {author} {\bibfnamefont {P.~J.~J.}\
  \bibnamefont {O'Malley}}, \bibinfo {author} {\bibfnamefont {D.}~\bibnamefont
  {Sank}}, \bibinfo {author} {\bibfnamefont {A.}~\bibnamefont {Vainsencher}},
  \bibinfo {author} {\bibfnamefont {H.}~\bibnamefont {Wang}}, \bibinfo {author}
  {\bibfnamefont {T.~C.}\ \bibnamefont {White}}, \bibinfo {author}
  {\bibfnamefont {A.~N.}\ \bibnamefont {Cleland}}, \ and\ \bibinfo {author}
  {\bibfnamefont {J.~M.}\ \bibnamefont {Martinis}},\ }\href {\doibase
  10.1103/PhysRevLett.110.150502} {\bibfield  {journal} {\bibinfo  {journal}
  {Phys. Rev. Lett.}\ }\textbf {\bibinfo {volume} {110}},\ \bibinfo {pages}
  {150502} (\bibinfo {year} {2013})}\BibitemShut {NoStop}%
\bibitem [{\citenamefont {Catelani}\ \emph {et~al.}(2011)\citenamefont
  {Catelani}, \citenamefont {Koch}, \citenamefont {Frunzio}, \citenamefont
  {Schoelkopf}, \citenamefont {Devoret},\ and\ \citenamefont
  {Glazman}}]{Catelani2011a}%
  \BibitemOpen
  \bibfield  {author} {\bibinfo {author} {\bibfnamefont {G.}~\bibnamefont
  {Catelani}}, \bibinfo {author} {\bibfnamefont {J.}~\bibnamefont {Koch}},
  \bibinfo {author} {\bibfnamefont {L.}~\bibnamefont {Frunzio}}, \bibinfo
  {author} {\bibfnamefont {R.}~\bibnamefont {Schoelkopf}}, \bibinfo {author}
  {\bibfnamefont {M.}~\bibnamefont {Devoret}}, \ and\ \bibinfo {author}
  {\bibfnamefont {L.}~\bibnamefont {Glazman}},\ }\href@noop {} {\bibfield
  {journal} {\bibinfo  {journal} {Phys. Rev. Lett.}\ }\textbf {\bibinfo
  {volume} {106}},\ \bibinfo {pages} {077002} (\bibinfo {year}
  {2011})}\BibitemShut {NoStop}%
\bibitem [{\citenamefont {Yan}\ \emph {et~al.}(2015)\citenamefont {Yan},
  \citenamefont {Gustavsson}, \citenamefont {Kamal}, \citenamefont {Birenbaum},
  \citenamefont {Sears}, \citenamefont {Hover}, \citenamefont {Gudmundsen},
  \citenamefont {Yoder}, \citenamefont {Orlando}, \citenamefont {Clarke},
  \citenamefont {Kerman},\ and\ \citenamefont {Oliver}}]{Yan2014a}%
  \BibitemOpen
  \bibfield  {author} {\bibinfo {author} {\bibfnamefont {F.}~\bibnamefont
  {Yan}}, \bibinfo {author} {\bibfnamefont {S.}~\bibnamefont {Gustavsson}},
  \bibinfo {author} {\bibfnamefont {A.}~\bibnamefont {Kamal}}, \bibinfo
  {author} {\bibfnamefont {J.}~\bibnamefont {Birenbaum}}, \bibinfo {author}
  {\bibfnamefont {A.~P.}\ \bibnamefont {Sears}}, \bibinfo {author}
  {\bibfnamefont {D.}~\bibnamefont {Hover}}, \bibinfo {author} {\bibfnamefont
  {T.~J.}\ \bibnamefont {Gudmundsen}}, \bibinfo {author} {\bibfnamefont
  {J.~L.}\ \bibnamefont {Yoder}}, \bibinfo {author} {\bibfnamefont {T.~P.}\
  \bibnamefont {Orlando}}, \bibinfo {author} {\bibfnamefont {J.}~\bibnamefont
  {Clarke}}, \bibinfo {author} {\bibfnamefont {A.~J.}\ \bibnamefont {Kerman}},
  \ and\ \bibinfo {author} {\bibfnamefont {W.~D.}\ \bibnamefont {Oliver}},\
  }\href@noop {} {\bibfield  {journal} {\bibinfo  {journal} {submitted for
  publication}\ } (\bibinfo {year} {2015})}\BibitemShut {NoStop}%
\bibitem [{\citenamefont {Ambegaokar}\ and\ \citenamefont
  {Baratoff}(1963{\natexlab{a}})}]{Ambegaokar63a}%
  \BibitemOpen
  \bibfield  {author} {\bibinfo {author} {\bibfnamefont {V.}~\bibnamefont
  {Ambegaokar}}\ and\ \bibinfo {author} {\bibfnamefont {A.}~\bibnamefont
  {Baratoff}},\ }\href@noop {} {\bibfield  {journal} {\bibinfo  {journal}
  {Phys. Rev. Lett.}\ }\textbf {\bibinfo {volume} {10}},\ \bibinfo {pages}
  {486} (\bibinfo {year} {1963}{\natexlab{a}})}\BibitemShut {NoStop}%
\bibitem [{\citenamefont {Ambegaokar}\ and\ \citenamefont
  {Baratoff}(1963{\natexlab{b}})}]{Ambegaokar63b}%
  \BibitemOpen
  \bibfield  {author} {\bibinfo {author} {\bibfnamefont {V.}~\bibnamefont
  {Ambegaokar}}\ and\ \bibinfo {author} {\bibfnamefont {A.}~\bibnamefont
  {Baratoff}},\ }\href@noop {} {\bibfield  {journal} {\bibinfo  {journal}
  {Phys. Rev. Lett.}\ }\textbf {\bibinfo {volume} {11}},\ \bibinfo {pages}
  {104} (\bibinfo {year} {1963}{\natexlab{b}})}\BibitemShut {NoStop}%
\end{thebibliography}%


\begin{thebibliography}{0}%
\makeatletter
\providecommand \@ifxundefined [1]{%
 \@ifx{#1\undefined}
}%
\providecommand \@ifnum [1]{%
 \ifnum #1\expandafter \@firstoftwo
 \else \expandafter \@secondoftwo
 \fi
}%
\providecommand \@ifx [1]{%
 \ifx #1\expandafter \@firstoftwo
 \else \expandafter \@secondoftwo
 \fi
}%
\providecommand \natexlab [1]{#1}%
\providecommand \enquote  [1]{``#1''}%
\providecommand \bibnamefont  [1]{#1}%
\providecommand \bibfnamefont [1]{#1}%
\providecommand \citenamefont [1]{#1}%
\providecommand \href@noop [0]{\@secondoftwo}%
\providecommand \href [0]{\begingroup \@sanitize@url \@href}%
\providecommand \@href[1]{\@@startlink{#1}\@@href}%
\providecommand \@@href[1]{\endgroup#1\@@endlink}%
\providecommand \@sanitize@url [0]{\catcode `\\12\catcode `\$12\catcode
  `\&12\catcode `\#12\catcode `\^12\catcode `\_12\catcode `\%12\relax}%
\providecommand \@@startlink[1]{}%
\providecommand \@@endlink[0]{}%
\providecommand \url  [0]{\begingroup\@sanitize@url \@url }%
\providecommand \@url [1]{\endgroup\@href {#1}{\urlprefix }}%
\providecommand \urlprefix  [0]{URL }%
\providecommand \Eprint [0]{\href }%
\providecommand \doibase [0]{http://dx.doi.org/}%
\providecommand \selectlanguage [0]{\@gobble}%
\providecommand \bibinfo  [0]{\@secondoftwo}%
\providecommand \bibfield  [0]{\@secondoftwo}%
\providecommand \translation [1]{[#1]}%
\providecommand \BibitemOpen [0]{}%
\providecommand \bibitemStop [0]{}%
\providecommand \bibitemNoStop [0]{.\EOS\space}%
\providecommand \EOS [0]{\spacefactor3000\relax}%
\providecommand \BibitemShut  [1]{\csname bibitem#1\endcsname}%
\let\auto@bib@innerbib\@empty
\end{thebibliography}%

\end{document}



\newcommand{\addSG}[1]{\textcolor{magenta}{#1}}
\newcommand{\mpar}[1]{\marginpar{\small \it \textcolor{magenta}{#1}}}
\newcommand*{\Psw}{\ensuremath{P_\mathrm{sw}}\xspace}
\newcommand*{\tpw}{\ensuremath{t_\mathrm{pw}}\xspace}
\newcommand*{\rotq}{\ensuremath{\theta}\xspace}
\newcommand*{\Aq}{\ensuremath{A_\mathrm{Q}}\xspace}
\newcommand*{\Ai}{\ensuremath{A_\mathrm{I}}\xspace}
\newcommand*{\Hint}{\ensuremath{\hat{H}_\mathrm{int}}\xspace}
\newcommand*{\Hsub}{\ensuremath{\hat{H}_\mathrm{sub}}\xspace}
\newcommand*{\Hqb}{\ensuremath{\hat{H}_\mathrm{qb}}\xspace}
\newcommand*{\Htls}{\ensuremath{\hat{H}_\mathrm{TLS}}\xspace}
\newcommand*{\siX}{\ensuremath{\hat{\sigma}_\mathrm{x}}\xspace}
\newcommand*{\siY}{\ensuremath{\hat{\sigma}_\mathrm{y}}\xspace}
\newcommand*{\siZ}{\ensuremath{\hat{\sigma}_\mathrm{z}}\xspace}
\newcommand*{\siXq}{\ensuremath{\hat{\sigma}_\mathrm{x}^\mathrm{qb}}\xspace}
\newcommand*{\siYq}{\ensuremath{\hat{\sigma}_\mathrm{y}^\mathrm{qb}}\xspace}
\newcommand*{\siZq}{\ensuremath{\hat{\sigma}_\mathrm{z}^\mathrm{qb}}\xspace}
\newcommand*{\siXt}{\ensuremath{\hat{\sigma}_\mathrm{x}^\mathrm{TLS}}\xspace}
\newcommand*{\siYt}{\ensuremath{\hat{\sigma}_\mathrm{y}^\mathrm{TLS}}\xspace}
\newcommand*{\siZt}{\ensuremath{\hat{\sigma}_\mathrm{z}^\mathrm{TLS}}\xspace}
\newcommand*{\siXs}{\ensuremath{\hat{\sigma}_\mathrm{x}^\mathrm{sub}}\xspace}
\newcommand*{\siYs}{\ensuremath{\hat{\sigma}_\mathrm{y}^\mathrm{sub}}\xspace}
\newcommand*{\siZs}{\ensuremath{\hat{\sigma}_\mathrm{z}^\mathrm{sub}}\xspace}
\newcommand*{\Isq}{\ensuremath{I_\mathrm{b}}\xspace}
\newcommand*{\PhiQ}{\ensuremath{\Phi_\mathrm{qb}}\xspace}
\newcommand*{\fqb}{\ensuremath{\omega_\mathrm{qb}}\xspace}
\newcommand*{\ftls}{\ensuremath{f_\mathrm{TLS}}\xspace}
\newcommand*{\fosc}{\ensuremath{f_\mathrm{osc}}\xspace}
\newcommand*{\dph}{\ensuremath{\delta\Phi}\xspace}
\newcommand*{\df}{\ensuremath{\delta\!f}\xspace}
\newcommand*{\dpint}{\ensuremath{\delta\Phi_\mathrm{int}}\xspace}
\newcommand*{\TphN}{\ensuremath{T_{\varphi,\mathrm{N}}}\xspace}
\newcommand*{\Tph}[1]{\ensuremath{T_{\varphi,\mathrm{#1}}}\xspace}
\newcommand*{\tp}{\ensuremath{t_\mathrm{p}}\xspace}
\newcommand*{\fRabi}{\ensuremath{f_\mathrm{Rabi}}\xspace}
\newcommand*{\Ian}{\ensuremath{I_\mathrm{antenna}^\mathrm{mw}}\xspace}
\newcommand*{\Imr}{\ensuremath{I_\mathrm{r}^\mathrm{mw}}\xspace}
\newcommand*{\emd}{\ensuremath{\varepsilon^\mathrm{mw}_\mathrm{direct}}\xspace}
\newcommand*{\emt}{\ensuremath{\varepsilon^\mathrm{mw}}\xspace}

\newcommand{\ket}[1]{\vert #1\rangle}
\newcommand{\bra}[1]{\langle #1\vert}
\newcommand{\braket}[1]{\langle #1\rangle}

\newcommand*{\PhiX}{\ensuremath{\Phi_\mathrm{X}}\xspace}
\newcommand*{\PhiZ}{\ensuremath{\Phi_\mathrm{Z}}\xspace}
\newcommand*{\fX}{\ensuremath{f_\mathrm{x}}\xspace}
\newcommand*{\fZ}{\ensuremath{f_\mathrm{z}}\xspace}
\newcommand*{\Ax}{\ensuremath{A_\mathrm{x}}\xspace}
\newcommand*{\Az}{\ensuremath{A_\mathrm{z}}\xspace}

\newcommand*{\TF}{\ensuremath{T_{\varphi F}}\xspace}
\newcommand*{\TE}{\ensuremath{T_{\varphi E}}\xspace}
\newcommand*{\GF}{\ensuremath{\Gamma_{\varphi F}}\xspace}
\newcommand*{\GE}{\ensuremath{\Gamma_{\varphi E}}\xspace}

\newcommand*{\GSs}{\ensuremath{\,\mathrm{GS/s}\xspace}}
\newcommand*{\mPh}{\ensuremath{\,\mathrm{m}\Phi_0}\xspace}
\newcommand*{\uPh}{\ensuremath{\,\mu\Phi_0}\xspace}

\newcommand*{\um}{\ensuremath{\,\mu\mathrm{m}}\xspace}
\newcommand*{\nm}{\ensuremath{\,\mathrm{nm}}\xspace}
\newcommand*{\mm}{\ensuremath{\,\mathrm{mm}}\xspace}
\newcommand*{\m}{\ensuremath{\,\mathrm{m}}\xspace}
\newcommand*{\sqm}{\ensuremath{\,\mathrm{m}^2}\xspace}
\newcommand*{\sqmm}{\ensuremath{\,\mathrm{mm}^2}\xspace}
\newcommand*{\squm}{\ensuremath{\,\mu\mathrm{m}^2}\xspace}
\newcommand*{\psqm}{\ensuremath{\,\mathrm{m}^{-2}}\xspace}
\newcommand*{\psqmV}{\ensuremath{\,\mathrm{m}^{-2}\mathrm{V}^{-1}}\xspace}
\newcommand*{\cm}{\ensuremath{\,\mathrm{cm}}\xspace}

\newcommand*{\nF}{\ensuremath{\,\mathrm{nF}}\xspace}
\newcommand*{\pF}{\ensuremath{\,\mathrm{pF}}\xspace}
\newcommand*{\pH}{\ensuremath{\,\mathrm{pH}}\xspace}

\newcommand*{\emob}{\ensuremath{\,\mathrm{m}^2/\mathrm{V}\mathrm{s}}\xspace}
\newcommand*{\edos}{\ensuremath{\,\mu\mathrm{C}/\mathrm{cm}^2}\xspace}
\newcommand*{\mbar}{\ensuremath{\,\mathrm{mbar}}\xspace}

\newcommand*{\A}{\ensuremath{\,\mathrm{A}}\xspace}
\newcommand*{\mA}{\ensuremath{\,\mathrm{mA}}\xspace}
\newcommand*{\nA}{\ensuremath{\,\mathrm{nA}}\xspace}
\newcommand*{\pA}{\ensuremath{\,\mathrm{pA}}\xspace}
\newcommand*{\fA}{\ensuremath{\,\mathrm{fA}}\xspace}
\newcommand*{\uA}{\ensuremath{\,\mu\mathrm{A}}\xspace}

\newcommand*{\Ohm}{\ensuremath{\,\Omega}\xspace}
\newcommand*{\kOhm}{\ensuremath{\,\mathrm{k}\Omega}\xspace}
\newcommand*{\MOhm}{\ensuremath{\,\mathrm{M}\Omega}\xspace}
\newcommand*{\GOhm}{\ensuremath{\,\mathrm{G}\Omega}\xspace}

\newcommand*{\Hz}{\ensuremath{\,\mathrm{Hz}}\xspace}
\newcommand*{\kHz}{\ensuremath{\,\mathrm{kHz}}\xspace}
\newcommand*{\MHz}{\ensuremath{\,\mathrm{MHz}}\xspace}
\newcommand*{\GHz}{\ensuremath{\,\mathrm{GHz}}\xspace}
\newcommand*{\THz}{\ensuremath{\,\mathrm{THz}}\xspace}

\newcommand*{\K}{\ensuremath{\,\mathrm{K}}\xspace}
\newcommand*{\mK}{\ensuremath{\,\mathrm{mK}}\xspace}

\newcommand*{\kV}{\ensuremath{\,\mathrm{kV}}\xspace}
\newcommand*{\V}{\ensuremath{\,\mathrm{V}}\xspace}
\newcommand*{\mV}{\ensuremath{\,\mathrm{mV}}\xspace}
\newcommand*{\uV}{\ensuremath{\,\mu\mathrm{V}}\xspace}
\newcommand*{\nV}{\ensuremath{\,\mathrm{nV}}\xspace}

\newcommand*{\eV}{\ensuremath{\,\mathrm{eV}}\xspace}
\newcommand*{\meV}{\ensuremath{\,\mathrm{meV}}\xspace}
\newcommand*{\ueV}{\ensuremath{\,\mu\mathrm{eV}}\xspace}

\newcommand*{\T}{\ensuremath{\,\mathrm{T}}\xspace}
\newcommand*{\mT}{\ensuremath{\,\mathrm{mT}}\xspace}
\newcommand*{\uT}{\ensuremath{\,\mu\mathrm{T}}\xspace}

\newcommand*{\dBm}{\ensuremath{\,\mathrm{dBm}}\xspace}

\newcommand*{\ms}{\ensuremath{\,\mathrm{ms}}\xspace}
\newcommand*{\s}{\ensuremath{\,\mathrm{s}}\xspace}
\newcommand*{\us}{\ensuremath{\,\mathrm{\mu s}}\xspace}
\newcommand*{\ns}{\ensuremath{\,\mathrm{ns}}\xspace}
\newcommand*{\rpm}{\ensuremath{\,\mathrm{rpm}}\xspace}
\newcommand*{\minute}{\ensuremath{\,\mathrm{min}}\xspace}
\newcommand*{\degree}{\ensuremath{\,^\circ\mathrm{C}}\xspace}

\renewcommand{\thefigure}{S\arabic{figure}}
\renewcommand{\thetable}{S\arabic{table}}
\newcommand*{\EqRef}[1]{Eq.\,\ref{#1}}
\newcommand*{\FigRef}[1]{Fig.\,\ref{#1}}
\newcommand*{\TabRef}[1]{Tab.\,S\ref{#1}}
\newcommand*{\dd}[2]{\mathrm{d}#1/\mathrm{d}#2}
\newcommand*{\ddf}[2]{\frac{\mathrm{\partial}#1}{\mathrm{\partial}#2}}
\newcommand*{\dddf}[2]{\frac{\mathrm{d}#1}{\mathrm{d}#2}}

\title{Supplementary Material to "Thermal and Residual Excited-State Population in a 3D Transmon Qubit"}


 \author{X. Y. Jin$^{1}$}
 \email{jin@mit.edu}
 \author{A. Kamal$^1$}
 \author{A. P. Sears$^{2}$}
 \author{T. Gudmundsen$^{2}$}
 \author{D. Hover$^{2}$}
 \author{J. Miloshi$^{2}$}
 \author{R.~Slattery$^{2}$}
 \author{F. Yan$^{1}$}
 \author{J. Yoder$^{2}$}
 \author{T. P. Orlando$^{1}$}
 \author{S. Gustavsson$^{1}$}
 \author{W. D. Oliver$^{1,2}$}
 \affiliation{
$^{1}$Research Laboratory of Electronics, Massachusetts Institute of Technology, Cambridge, MA 02139, USA \\
$^{2}$MIT Lincoln Laboratory, 244 Wood Street, Lexington, MA 02420, USA
}

\date{\today}
\maketitle

\section{Supplementary Material}

\subsection{Experimental set-up}

\begin{figure}[b]
    \includegraphics[width=8cm]{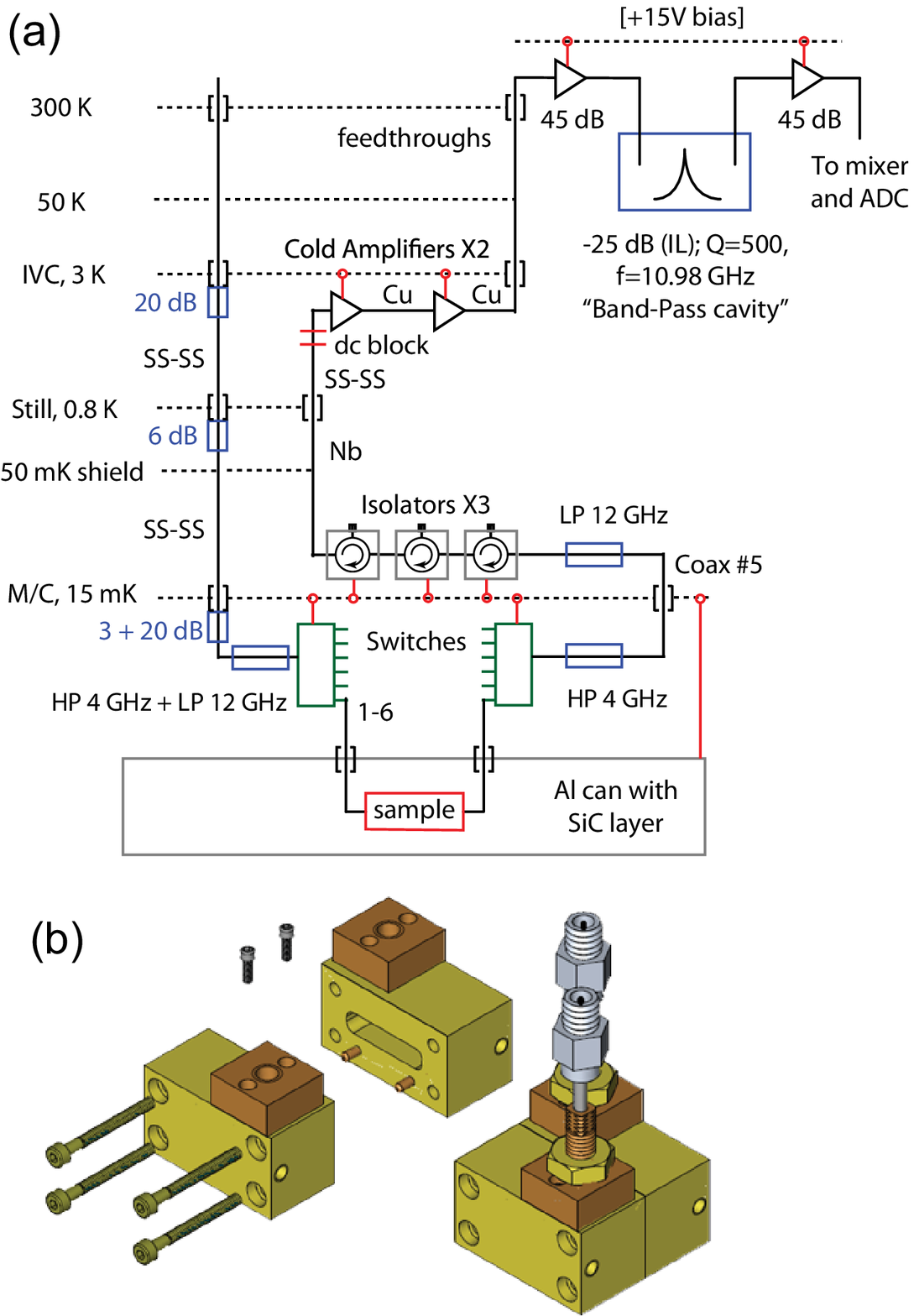}
    \caption{(a). The schematics of the dilution refrigerator and part of the setup. After the room temperature amplifiers, the signal is mixed down to 25 MHz and is sent to an ADC card. (b). The cavity design. Note that the coax pins are equipped with threaded shell, which are screwed and fixed onto the microwave ports. One can tune the coupling Q by rotating the coax pin against the holder.}
    \label{Fig:supp_set-up}
\end{figure}

Experiments were carried out in a cryogen-free dilution refrigerator from Leiden Cryogenics (model CF-450) with a base temperature of 15 mK. A Lakeshore 370 temperature controller monitors the mixing chamber temperature using a calibrated Speer thermometer mounted close to the qubit. The Lakeshore also sets the mixing chamber temperature (0.1 mK resolution) through PID feedback control of a current sent to a 100-Ohm resistor (heater) mounted on the mixing chamber. The heater is located far from the device and the thermometer.

The device is isolated from higher temperature stages of the refrigerator through a series of nested shields. Starting with the qubit (and working outwards), the device is mounted inside an aluminum cavity (Fig.~\ref{Fig:supp_set-up}). This cavity is thermally anchored to a copper cold finger attached to the mixing chamber plate with variable temperature down to 15 mK. The cavity is located inside an aluminum box with feedthroughs allowing the coaxial cables and the copper finger to enter it, and it is also anchored to the copper cold finger. The inside of the aluminum box is lined with a thin layer of copper (also anchored to the cold finger), and the copper is coated with SiC powder (20 grit) and Stycast 2850FT~\cite{Persky99a,UCSB_Wiring}. We note that $T_1$ times do not appear to vary significantly whether the device is mounted inside or outside this box, but we have not performed a systematic study. We also note that devices measured previously in an aluminum can without the microwave absorbing material had similarly low effective temperatures. Next, a $\mu$-metal can is attached to the mixing chamber plate. Then, there are two brass cans: one at the 50-mK ``cold-plate'' stage and one at the 800-mK still stage. Finally, there is a hermetically sealed inner-vacuum-chamber (IVC) can at the 3-K stage. Care is taken to block line-of-sight paths between temperature stages.

On the input side, there is 49 dB attenuation from discrete attenuators (XMA) mounted between the 3-K stage and the mixing chamber. There is additional frequency-dependent loss from the coaxial cables. Two microwave switches (Radiall R591762600) mounted on the cold finger allow up to six samples to share the same input and output lines. On the input and output sides of the device, a high-pass (RLC F-18948, 4 GHz cutoff) and low-pass (RLC L-3615, 12.4 GHz cutoff) filter provide a net 4-12.4 GHz passband with $> 60$ dB isolation in the high-frequency stop band out to at least 40 GHz. After the output filters, 3 isolators (Quinstar / Pamtech, model CWJ1019KS414, 3-12 GHz, with approximately 15-20 dB isolation each) are mounted on the mixing chamber. The signal is amplified using JPL/Caltech cryogenic preamplifiers (1-12 GHz, 30 dB gain, 4-6K noise temperature). We attempted to improve SNR by combining the two amplifiers in series for the 3D transmon data. In the series case, the gain of the second preamplifier is largely saturated by the output of the first preamplifier, but there is otherwise no adverse effect. These are followed by room-temperature amplifiers (MITEQ, AMF-5D-00101200-23-10P, 0.1-12 GHz, 43 dB gain, 145-200K noise temperature). Between the two room temperature amplifies, a 3D cavity (loaded with a sapphire chip) is used as a narrow-band band-pass filter to improve the signal-to-noise ratio and to reduce the saturation of the second amplifier. Nonetheless, the second amplifier was partially saturated with no adverse effect. The signal was then demodulated and digitized (Acquiris U1084A). In subsequent measurements with a flux qubit (see below), it was determined that the use of series amplifiers (cold and warm) did not produce a substantial SNR improvement due to amplifier saturation.

The resonator cavity was machined from Aluminum 6061-T6, and its size is $19 \times 17 \times 4$ mm$^3$. The internal $Q$ of the cavity is measured to be above $10^6$. Microwave ports are located at opposing corners of the cavity, where the electric field is relatively weak. The coupling pins are made of semi-rigid coax cables with 15 mm of outer conductor removed and inner conductor exposed. The coupling pin is equipped with a fine-pitch threaded shell, which is threaded onto the cavity and fixed with a nut. The coupling Q is tuned precisely by rotating the coupling pin. The cavity was sealed with Indium wire to make it light-tight against the residual thermal photons at the mixing chamber stage. However, this is likely unnecessary; all other 2D qubits had no similar precautions, yet achieved similarly low effective temperatures.

\subsection*{Device parameters}
Our single-junction superconducting transmon qubit features a Josephson energy $E_{\textrm{J}} = h \times 14.07$ GHz and charging energy $E_{\textrm{C}} = h \times 0.24$ GHz~\cite{Peterer2015a}, where $h$ is Planck's constant.
The ratio $E_{\textrm{J}}/E_{\textrm{C}} = 58$ places the qubit in the charge-insensitive transmon regime with transition frequencies $f_{\rm ge}=4.97$ GHz and $f_{\rm ef}=4.70$ GHz.
The qubit is controlled using a circuit-QED approach through its strong dispersive coupling ($g / 2\pi = 160$ MHz) to an aluminum cavity with a TE101 mode frequency of 10.976 GHz (when loaded with a sapphire chip), an internal quality factor $Q_{\mathrm{i}} > 10^6$, and two ports with a net coupling $Q_{\mathrm{c}} = 10^5$.
In estimating the quasiparticle relaxation rate, we take $2\Delta = 340$ $\mu$eV as the superconducting energy gap for aluminum.
The capacitance $C=80$ fF is derived from the charging energy $E_{\textrm{C}} = e^2/2C$.
The normal-state resistance $R_{\textrm{N}}=9.5$ k$\Ohm$ is found from the Ambegaokar-Baratoff relation $R_{\textrm{N}} = (\pi / 4) (2 \Delta / I_{\textrm{c}})$~\cite{Ambegaokar63a}, where $I_{\textrm{c}}$ is the transmon junction critical current found from the Josephson energy $I_{\textrm{c}} \Phi_0 / 2 \pi$ with $\Phi_0 = h/2e$ the superconducting quantum unit of flux.

\subsection*{Protocol used to reduce low-frequency drift}

To address the problem of low-frequency drift of the signal during long averaging periods, the data were acquired in a cyclical manner.
Each cycle comprised four ``points'' taken in the order R1-R2-S1-S2, with each point comprising $A = 5 \times 10^3$ averages.
Therefore, each such cycle represents a total of $4A = 2 \times 10^4$ trials and required approximately 10 seconds acquisition time.
Using this method, low-frequency noise below approximately 0.1 Hz was experimentally mitigated during data acquisition.
In the absence of low-frequency drift, the estimators derived from each cycle are treated as samples and are assumed to be derived from independent and identically distributed, ergodic stochastic processes.

\subsection*{Repetition period and $T_1$ relaxation time}
Our experiment was performed with a 2 kHz repetition rate, corresponding to a period 500 $\mu$s. This period is in general somewhat short for a qubit with exponential relaxation time $T_1 = 80$ $\mu$s, and it could in principle be a source of error for a qubit with large excited state population. For example, if the qubit were prepared with 100\% excited state population, we would have a residual value 0.2\% after 500 $\mu$s. In our calibration experiment (main text, Figure 2), however, we operated in the range 0.2-5\% excited-state population. Even for the worst-case 5\% level, the residual drops to around 0.01\% after 500 $\mu$s, much less than the 0.067\% inferred from that experiment. In the direct measurement (Figure 3), the maximum excited-state population was even lower ($< 2\%$). At the saturation level $0.1\%$, the excited-state population error due to the repetition period was a negligible 0.005\%.

\subsection*{Data processing and error evaluation}

For each cycle described above, we calculate an estimator for the excited state population $\Pe^{(i)}$ (a sample value for the $i$-th cycle)
\begin{equation}
    \Pe^{(i)} = \frac{A_{sig}^{(i)}}{A_{sig}^{(i)} + A_{ref}^{(i)}},
\end{equation}
where
\begin{align}
    A_{\rm sig}^{(i)} &= S_2^{(i)} - S_1^{(i)} \\
    A_{\rm ref}^{(i)} &= R_2^{(i)} - R_1^{(i)}.
\end{align}
Note that although cycle $i$ require a total of $4A$ trials, each random variable in the estimator results from $A$ averages.
The cycles are repeated $C$ times to reach a total number of averages $N = CA$ sufficient to achieve a desired measurement resolution.
The reported excited state population $\Pe$ is then the sample mean $\overline{\Pe}$ over the $C$ samples (estimators) $\Pe^{(i)}$,
\begin{equation}
    \overline{\Pe} =\frac{1}{C}\sum_{i=1}^C \Pe^{(i)}.
\end{equation}
For example, the reported $\Pe=\overline{\Pe}=0.055\%$ at 15 mK resulted from a total of $N=1.5 \times 10^7$ averages comprising $C=3 \times 10^3$ samples (estimates).

The standard deviation $\sigma_{C}$ of the {\em sample distribution} is
\begin{equation}
    \sigma_{C} = \sqrt{\frac{1}{C} \sum_{i=1}^{C} (\Pe^{(i)} - \overline{\Pe})^2}.
\end{equation}
Note that $\sigma_{C}$ depends on the number of averages $A$. If we define $\sigma_{t}$ to be the standard deviation of the relevant stochastic processes impacting a single trial $t$, then we can assign $\sigma_{C} = \sigma_{t} / \sqrt{A}$. Obviously, more averaging per cycle will decrease $\sigma_{C}$, but it also may increase our sensitivity to low frequency noise.

The error bar on the estimate for $\Pe$ is the standard error of the mean,
\begin{equation}
    \delta\Pe = \frac{\sigma_C}{\sqrt{C}},
\end{equation}
where we have assumed independent and identically distributed samples. For example, at 15 mK the standard error $\delta \Pe=0.039\%$ was obtained for the sample mean $\overline{\Pe}=0.055\%$. That is, we have $\Pe=0.055 \pm 0.039 \%$ at bath temperature of 15 mK.

In Fig.~\ref{Fig:histogram}, we plot histograms of the samples $\Pe^i$ using a fixed $0.5 \%$ bin-width for temperatures $T=15 \ldots 60$ mK. We tabulate in Table~\ref{tab:stats} the results from the arithmetic data processing as well as from Gaussian fitting to the sample data.

We have presented a treatment based on statistical sampling of the estimator $\Pe^{(i)}$ defined in Eq.~1. This analysis was sufficient for our goal in this work, namely, to measure the excited state population versus temperature. A further step would be to define the error propagation from the individual random variables $S_1$, $S_2$, $R_1$ and $R_2$ in Eqs.~2 and~3 into the estimator for $\Pe^{(i)}$. This approach is also of potential interest, because it takes as its starting point the underlying noise processes of the measured quantities (as opposed to our defined estimator). In principle, the two approaches would lead to the same measured noise (e.g., as predicted in Eq.~5).

\subsection{Addressing the potential for skewness of the sample distribution}
The excited-state population of the qubit is never negative. Based on this fact, one might expect to observe (in principle) a degree of skewness in the sample data. However, our sample data is clearly Gaussian distributed with minimal skewness (see Fig.~\ref{Fig:histogram}). This arises from the fact that our dominant noise source is not related to the thermodynamics of the excited state population (an ``intrinsic'' stochastic process),
but rather are limited by the external noise associated with our measurement chain such as amplifier noise (``extrinsic'' stochastic processes). The extrinsic noise is predominantly Gaussian-distributed, zero-mean, and white over the measurement bandwidth. In the limit of small qubit population, the population is proportional to the voltage difference $S2-S1$, and, when the external voltage noise in the setup is comparable to $S2-S1$, we expect the measured signal to sometimes drops below zero due to stochastic voltage fluctuations.  However, upon averaging sufficiently and collecting the sample data in a histogram as in Fig.~\ref{Fig:histogram} and Fig.~\ref{Fig:C-shunt}(b), it is clear that the mean value over many averages is greater than zero.

Consequently, the histograms in Fig.~\ref{Fig:histogram} are very well fit to Gaussian functions and the sample data have little skewness. Sample data in the negative tails of the Gaussians are due solely to extrinsic noise. In Table~\ref{tab:stats}, we list the calculated moment-coefficient of skewness $\gamma_1 \equiv \mu_3 / \mu_2^{3/2}$, where with $\mu_i$ is the $i$th central moment, for for temperatures $T=15 \ldots 60$ mK. In all cases, the skewness is negligible.
\vspace{0.2in}

\subsection{Effective temperature of a capacitively shunted flux qubit}
We have also measured the residual excited-state population of a capacitively-shunted flux qubit measured using dispersive readout via a coplanar waveguide cavity~\cite{Yan2014a}. The qubit frequency is 4.7 GHz, and it is capacitively coupled to a coplanar waveguide-type resonator with $f_\mathrm{resonator}=8.3~\mathrm{GHz}$, $Q_c = 5000$, and with a coupling strength $g = 100~\mathrm{MHz}$.
%
In this experiment, only a single cryogenic amplifier and a single MITEQ amplifier were used (cf., Fig.~\ref{Fig:supp_set-up}). The band-pass cavity was also removed. Following the MITEQ amplifier, an SRS low-frequency preamplifier (SR445A, DC - 350 MHz bandwidth, 5 V/V gain) was added before the digitizer.
Fig.~\ref{Fig:C-shunt}(a) shows the excited-state population, measured repeatedly for 16 hours ($A=3 \times 10^5$, $C=10^2$, $N=3 \times 10^7$) using the techniques described in the main text. In Fig.~\ref{Fig:C-shunt}(b), we plot a histogram of the data in Fig.~\ref{Fig:C-shunt}(a). The sample statistics include $\Pe = 0.17\%$, $\sigma_C = 0.097\%$, and $\delta \Pe = 0.0097\%$. Therefore, the average population is
0.17\% +/- 0.0097\%, which for a qubit frequency of 4.7 GHz corresponds to an effective temperature of 35 mK. 

\begin{figure*}[t]
 \includegraphics[width=15cm]{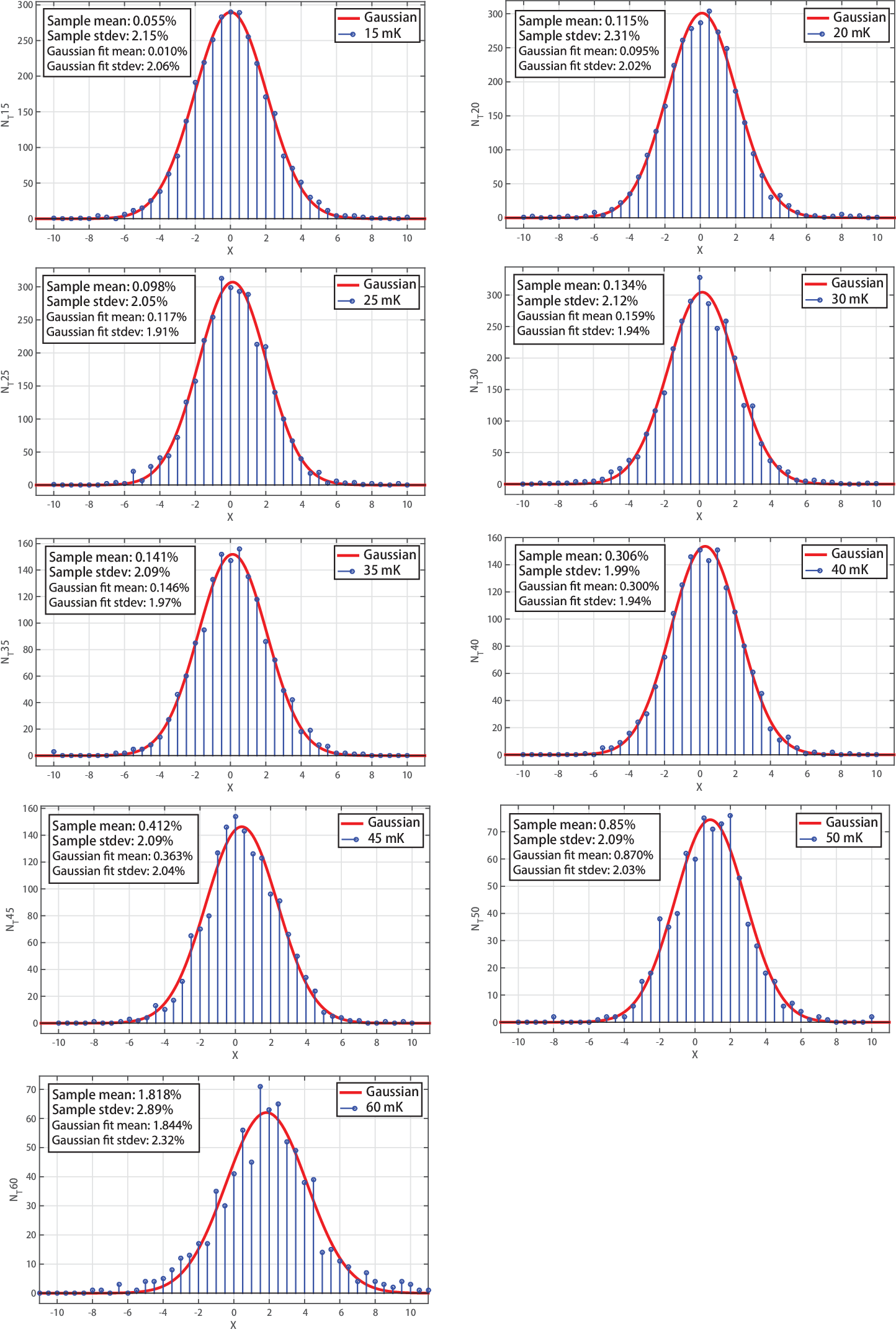}
    \caption{Histograms of sample data for temperatures $T=15\ldots60$ mK. The red solid line is a fit to a Gaussian function. Insets show sample statistics and Gaussian fit parameters.}
    \label{Fig:histogram}
\end{figure*}

\begin{figure*}[t]
    \includegraphics[width=15cm]{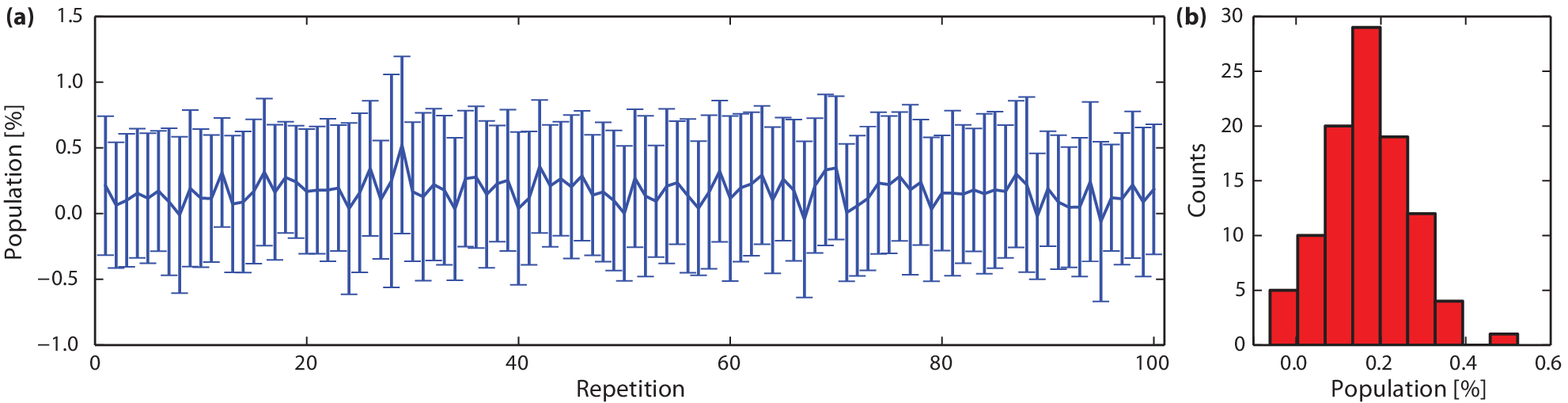}
    \caption{
    (a) Excited-state population in a capacitively shunted flux qubit. Each data point corresponds to about 10 minutes of averaging, and the total measurement time is approximately 16 hours.
    (b) Population data from panel (a), collected in a histogram.  The average population is $\Pe = 0.17\%$, the Gaussian standard deviation is 0.097\%, and the standard error is $\delta \Pe = 0.0097\%$. }
    \label{Fig:C-shunt}
\end{figure*}

As in the previous section, note that there are data points in Fig.~\ref{Fig:C-shunt}(a) with population below zero, and that the histogram in Fig.~\ref{Fig:C-shunt}(b) extends into the negative-population region.  The degree to which this happens is less than in Fig.~\ref{Fig:histogram}, because the number of averages per sample is much larger ($A=3 \times 10^5$) and the extrinsic noise level is lower due to the measurement-chain rearrangement.
Moreover, as before, the distribution of points is symmetric around the mean value, as expected for independent and symmetric noise sources such as voltage fluctuations of the amplifiers in the system.  The skewness $\gamma_1= -0.04 \pm 0.04$ in Fig.~\ref{Fig:C-shunt}(b) is close to zero.

\begin{table*}[b]
    \caption{Selected sample statistics and Gaussian fit parameters for data in Fig.~\ref{Fig:histogram}. $A$ is the number of averages per point, $C$ is the number of cycles (samples), $N$ is the total number of averages, $\Pe$ is the reported sample mean, $\sigma_C$ is the standard deviation of the samples, $\delta \Pe$ is the standard error of the mean (error bar) $\Pe$, and $\gamma_1$ is the skewness. \vspace{0.2in} }
    \centering 
    \setlength{\tabcolsep}{7pt}
    \begin{tabular}{c c | r r r r r r r | r r} 
        \hline\hline 
        \multicolumn{2}{c}{Temperature (mK)} & \multicolumn{7}{c}{Sample Statistics} & \multicolumn{2}{c}{Gaussian Fit} \\ [0.5ex]
        \multicolumn{1}{c}{Bath} & \multicolumn{1}{c}{$T_{\mathrm{eff}}$} & \multicolumn{1}{c}{$A$} & \multicolumn{1}{c}{$C$} & \multicolumn{1}{c}{$N$} & \multicolumn{1}{c}{$\sigma_C$} & \multicolumn{1}{c}{$\Pe$} & \multicolumn{1}{c}{$\delta \Pe$} & \multicolumn{1}{c}{$\gamma_1$} & \multicolumn{1}{c}{stdev} & \multicolumn{1}{c}{mean}  \\ [0.5ex]
        \hline 
        15 & $31.8 \pm (2.5,4.5)$ & $5 \times 10^3$ & $3.0 \times 10^3$ & $1.50 \times 10^7$ & 2.15\% & 0.055\% & 0.039\% & 0.081  & 2.04\% & 0.011\%  \\
        20 & $35.2 \pm (1.7,2.2)$ & $5 \times 10^3$ & $3.0 \times 10^3$ & $1.50 \times 10^7$ & 2.31\% & 0.115\% & 0.042\% & 0.102  & 1.97\% & 0.093\%  \\
        25 & $34.5 \pm (1.7,2.2)$ & $5 \times 10^3$ & $3.0 \times 10^3$ & $1.50 \times 10^7$ & 2.05\% & 0.098\% & 0.037\% & 0.000  & 1.92\% & 0.122\%  \\
        30 & $36.1 \pm (1.4,1.8)$ & $5 \times 10^3$ & $3.0 \times 10^3$ & $1.50 \times 10^7$ & 2.12\% & 0.134\% & 0.039\% & -0.056 & 1.93\% & 0.158\%  \\
        35 & $36.3 \pm (1.9,2.5)$ & $5 \times 10^3$ & $1.5 \times 10^3$ & $7.50 \times 10^6$  & 2.09\% & 0.141\% & 0.054\% & -0.272 & 1.94\% & 0.120\%  \\
        40 & $41.2 \pm (1.1,1.3)$ & $5 \times 10^3$ & $1.5 \times 10^3$ & $7.50 \times 10^6$  & 1.99\% & 0.306\% & 0.051\% & 0.026  & 1.94\% & 0.304\%  \\
        45 & $43.4 \pm (1.0,1.1)$ & $5 \times 10^3$ & $1.5 \times 10^3$ & $7.50 \times 10^6$  & 2.09\% & 0.412\% & 0.054\% & 0.045  & 2.04\% & 0.363\%  \\
        50 & $50.0 \pm (0.9,1.0)$ & $5 \times 10^3$ & $7.5 \times 10^2$  & $3.75 \times 10^6$  & 2.09\% & 0.850\% & 0.076\% & 0.055  & 1.99\% & 0.865\%  \\
        60 & $59.5 \pm (0.9,0.9)$ & $5 \times 10^3$ & $7.5 \times 10^2$  & $3.75 \times 10^6$  & 2.89\% & 1.818\% & 0.106\% & 0.222  & 2.29\% & 1.850\%  \\
        [1ex] 
        \hline 
    \end{tabular}
    \label{tab:stats}
\end{table*}
